# Mesoscale properties of protein clusters determine the size and nature of liquid-liquid phase separation (LLPS)

Gonen Golani[1,2], Manas Seal[2], Mrityunjoy Kar[3,4], Anthony A. Hyman[3], Daniella Goldfarb[2] and Samuel Safran[2]

[1] Department of Physics, University of Haifa, 3498838, Haifa, Israel.

[2] Department of Chemical and Biological Physics, Weizmann Institute of Science, Rehovot, Israel.

[3] Max Planck Institute of Cell Biology and Genetics, Dresden, Germany

[4] Institute of Biofunctional Polymer Materials, Leibniz Institute of Polymer Research Dresden, Dresden, Germany

Emails: ggolani@univ.haifa.ac.il, sam.safran@weizmann.ac.il

## Abstract

The observation of Liquid-Liquid Phase Separation (LLPS) in biological cells has dramatically shifted the paradigm that soluble proteins are uniformly dispersed in the cytoplasm or nucleoplasm. The LLPS region is preceded by a one-phase solution, where recent experiments have identified clusters in an aqueous solution with $10^2$-$10^3$ proteins. Here, we theoretically consider a core-shell model with mesoscale core, surface, and bending properties of the clusters' shell and contrast two experimental paradigms for the measured cluster size distributions of the Cytoplasmic Polyadenylation Element Binding-4 (CPEB4) and Fused in Sarcoma (FUS) proteins. The fits to the theoretical model and earlier electron paramagnetic resonance (EPR) experiments suggest that the same protein may exhibit hydrophilic, hydrophobic, and amphiphilic conformations, which act to stabilize the clusters. We find that CPEB4 clusters are much more stable compared to FUS clusters, which are less energetically favorable. This suggests that in CPEB4, LLPS consists of large-scale aggregates of clusters, while for FUS, clusters coalesce to form micron-scale LLPS domains.

**Introduction**

In recent years, a paradigm shift has transformed our understanding of cellular organization, with the discovery that many proteins undergo liquid-liquid phase separation (LLPS) to form condensates. This challenges the long-held view of cellular components as uniformly dispersed or structurally confined (e.g., the cytoskeleton, chromatin) [1]. The LLPS of proteins is generally attributed to their stronger mutual attraction compared to their affinity for the cell's aqueous environment [2]. The physics of these interactions and the resulting condensates are often quantified in controlled, *in-vitro* studies [1, 3-6] with a small number of components to avoid the highly realistic but complex nature of the dynamic, active, and multi-component cellular environment [7-10].

Beyond LLPS, it has been observed that certain proteins with disordered domains can form large (30–300 nm) self-assembled clusters even within the single-phase regime where LLPS is absent [4, 5, 11-13]. The observation of large mesoscale clusters is puzzling since small molecules in aqueous solution form clusters or oligomers with quite low probability in the one-phase region. In contrast, these protein clusters contain hundreds or thousands of solute monomers [4, 5, 11]. In-vitro studies have characterized this phenomenon, revealing distinct behaviors as a function of protein concentration. We focus on two representative cases: $CPEB4_{NTD}$ (the intrinsically disordered N-terminal domain of Cytoplasmic Polyadenylation Element Binding protein 4, denoted as CPEB4), whose cluster size is only weakly dependent on concentration [11], and FUS, whose cluster size shows a strong dependence on concentration [4].

Previous cluster formation models have been based on the sticker-spacer framework for intrinsically disordered proteins [4, 14, 15]. Molecular dynamics simulations using this framework predicted that proteins at the cluster surface adopt more elongated conformations and align predominantly in the normal direction to the cluster's surface [15]. These models also suggested that the sticker-sticker interaction energy required to form the observed large clusters is in the range of ~10 $k_BT$ [4]. A core-shell model was proposed for FUS in coarse-grained simulations, where the proteins were modeled as a diblock copolymer, revealing an internal core-shell structure [16]. Core-shell clusters were computationally simulated in peptide chains containing leucine and serine. In these simulations, the hydrophobic leucine residues preferentially occupied the core, while the hydrophilic serine residues were more likely to be found in the shell [17]. However, these computational models neither dealt with the distinct size dependence of the different types of clusters nor addressed the specific protein properties that allow cluster formation.

We suggest that the formation of these clusters is possible since disordered proteins, in contrast to small molecules undergoing LLPS, can have several different conformations. Some stabilize the cluster core, while others stabilize the cluster surface (shell). We quantify this using a generic mesoscale core–shell model and compare the predicted cluster size distribution with experimental data to extract the core and shell energies. The mesoscale nature of these energies distinguishes the clusters from relatively small oligomers whose properties are very sensitive to molecular details [18, 19]. In contrast to molecular simulations [4, 14-17], the mesoscale nature of our approach bypasses the detailed molecular calculation of protein conformations. Instead, we characterize those conformations by the contributions to the energies relevant to cluster self-assembly using only a few parameters extracted from the experiment.

We used a similar approach to investigate the temperature dependence of the core and shell energies of CPEB4 clusters as they approach LLPS [11]. In this work, we extend the model to derive a scaling law for the dependence of the clusters' mean size on the total protein concentration [20], as well as the size distribution around the mean value. This concentration dependence provides estimates for the bending energies of the clusters, as well as convincing evidence confirming that FUS clusters that precede LLPS lie below their Critical Aggregation Concentration [21] (CAC). In contrast, CPEB4 clusters lie above their CAC [22]. Further, we compare the energetic parameters obtained from fits of the experiment to the theory for Ddx4n1 and α-synuclein clusters [13].

In summary, our comparison of experiment and theory leads us to conclude that these proteins (CPEB4 vs FUS and the others which we show behave like FUS such as Ddx4n1, and α-synuclein) represent two different paradigms for cluster formation, leading to LLPS: CPEB4 clusters are more energetically stable and favorable than those in FUS, which are essentially large concentration fluctuations. The cluster energetics and stability have important implications for the nature of the large, micron-scale domains in the LLPS region of the phase diagram. For CPEB4, the analysis, as well as EPR measurements [5, 11], strongly suggest that LLPS is induced by aggregation of relatively stable clusters. In contrast, the less stable FUS clusters likely coalesce in the LLPS phase, resulting in large domains with proteins in their hydrophobic conformation.

**Methods**

*Theoretical Model of Cluster Structure and Protein Configurations*

The theoretical model aims to describe the cluster size distribution within the framework of continuum mesoscopic theory, drawing parallels to other mesoscale structures observed in self-assembling amphiphiles. Specifically, we *propose* that these clusters resemble microemulsions (swollen micelles), where an amphiphilic surface layer separates a hydrophobic core from the surrounding aqueous solvent [21, 22]. We show below that this is consistent with the measured, mesoscale cluster equilibrium distribution and the concentration dependence of the average cluster size. The distinct behavior of these quantities leads us to delineate two paradigms (CPEB4 with low-energy clusters and FUS with high-energy clusters) for cluster formation and their relationship to LLPS; these are the main conclusions of our paper.

While traditional microemulsions require "oil" and "amphiphiles," the protein-water-salt solutions of interest here achieve similar behavior without these additives. Instead, the disordered protein domains, composed of sequences with both hydrophobic and hydrophilic regions, enable the same protein species to adopt distinct configurations based on their local environment. This adaptability gives rise to different ensembles of protein states in the dilute aqueous phase, the dense cluster core, and the interfacial cluster shell.

The clusters are stabilized when protein configurations in the shell minimize the interfacial energy cost. This stabilization is achieved when proteins in the shell behave like amphiphiles, adopting configurations akin to the role of surfactants in microemulsions, in which their hydrophobic amino acids are oriented (but still internally disordered as for block copolymers [23]) toward the dense core. In contrast, their hydrophilic amino acids face the aqueous phase. Molecular-scale evidence for this is provided in Supplementary Table S1 of the Supplementary Information (SI) and its accompanying explanatory caption.

Based on this, along with the aforementioned computational and experimental evidence, we hypothesize that proteins adopt three distinct configurations corresponding to the different regions of the system, as illustrated in Fig. 1:

1. **Water-soluble Configuration:** In the dilute aqueous solution, hydrophobic amino acids are collapsed into a "blob" (in the sense used in polymer physics [24]) shielded by the hydrophilic ones. In this configuration, interactions between amino acids within the same protein chain are in a minimal energy configuration, although disordered.
2. **Hydrophobic Configuration:** In the dense cluster core, hydrophobic amino acids are exposed and interact with those of other proteins, while the hydrophilic blobs are collapsed, driving cluster formation. We suggest that the hydrophobic blob maximizes the interaction with the neighboring proteins' hydrophobic blobs at the expense of water, which is partially depleted.
3. **Amphiphilic Configuration:** At the cluster surface, proteins orient their hydrophobic amino acids toward the core (still disordered as in a polymer blob) and their hydrophilic amino acids toward the aqueous phase (also disordered as in a blob). In this configuration, proteins are more extended due to their two-blob nature, compared to the previous cases [15]. Although a detailed molecular analysis, such as determining the size of hydrophobic and hydrophilic regions or their relative solubility energies, is beyond the scope of this work, we note that both CPEB4 and the FUS protein family (including hnRNPA3, EWSR1, and TAF15) can be modeled as comprising two distinct blobs: a relatively ordered RNA-binding domain and one or two intrinsically disordered tails (Fig. S1A). These two regions display distinct hydrophobicity indices [25] (Table S1), supporting the hypothesis that proteins within the cluster shell can behave like amphiphiles.

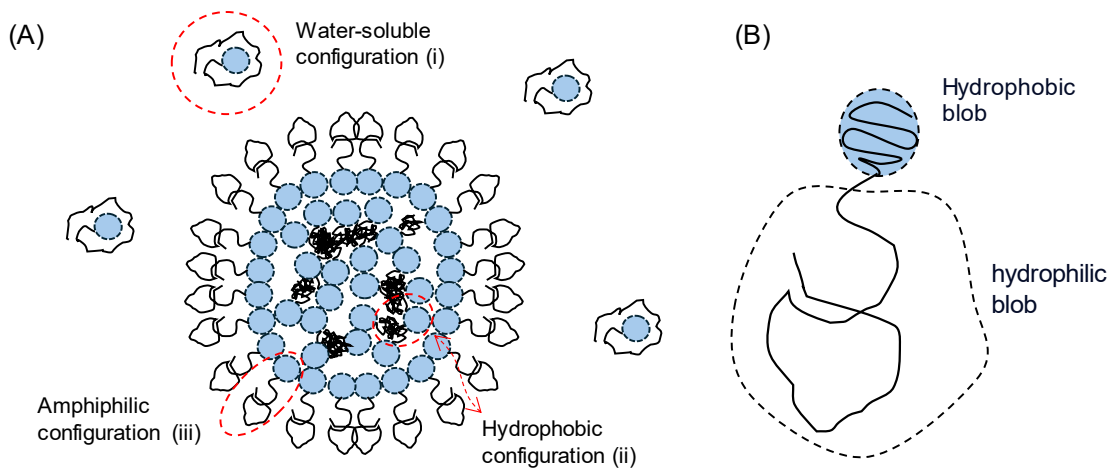

**Figure 1 – Protein Conformations. (A)** Illustration of conformations of dilute protein monomers in the solvent and those packed in the suggested core-shell structure of the clusters. The blue color denotes more hydrophobic parts, but does not imply that these segments are folded. The black, hydrophilic domain of the proteins in the cluster core is more collapsed. **(B)** The interfacial layer consists of more hydrophilic protein segments that face the solvent and more hydrophobic segments that face the core.

Direct experimental evidence for the coexistence of the three conformations comes from EPR measurements of the molecular rotational diffusion times of spin-labeled CPEB4 in which *three* distinct times, separated by two orders of magnitude, were observed [5, 11]. The shortest time is identified with the soluble conformation, the longest with the amphiphilic configuration, and an intermediate time with the proteins in the core. These are all the same chemical species, which can assume different conformations at a given temperature and concentration due to their disordered nature. Three typical sets of conformations were also shown using different molecular dynamics approaches in other proteins with intrinsically disordered domains [15, 26], suggesting that proteins at the surface of condensates assume different configurations than those at their core [14, 15].

*Theoretical model for the formation energies of the core and shell*

To account for the self-assembly energy of the clusters, our mesoscopic model considers the energy of the cluster core and shell. The cluster geometry is taken to be spherical, consistent with atomic force microscopy imaging [11] and electron microscopy [27] of CPEB4, and transmission electron microscopy of FUS clusters [4] in the salinity and temperature regime studied.

We note that long (100-200 nm) FUS fiber-like structures have been reported under different experimental conditions [28], which are outside the scope of the present work. The formation of the clusters we consider here is reversible upon change of temperature (CPEB4 [11]), dilution (α-synuclein [13]), and protein concentration (FUS [4]). Therefore, the theoretical model we developed here assumes the system is in *equilibrium*.

The two-blob amino acid structure of the proteins and the computationally predicted elongated structure at the cluster shell [15] indicate a nematic-like ordering of the cluster shell with non-zero bending modulus [22]. Therefore, we model the surface energy of the cluster in the spirit of an amphiphilic layer with interfacial tension, $\gamma$, bending modulus $\bar{\kappa}$, and spontaneous curvature $J_s$ [21]. For a spherical cluster, this results in the following surface energy [11, 22, 29]:

$$U_{shell} = 4\pi\gamma R^2 - 8\pi\kappa_B J_s R + 8\pi\bar{\kappa}. \qquad (1)$$

where $\bar{\kappa} \equiv \kappa_B + \frac{\kappa_G}{2}$, since for spherical clusters, the contributions of the bending, $\kappa_B$, and saddle-splay moduli, $\kappa_G$, cannot be distinguished.

The core free energy relative to that of a protein molecule in the solution is given by the product of the number of proteins in the cluster, $N$, and the difference between solubility energy, $\epsilon_B$, and the protein chemical potential, $\mu$,

$$U_{core} = N \cdot (\epsilon_B - \mu). \qquad (2)$$

The chemical potential of a protein molecule in a dilute solution (all the experimental systems considered here are very dilute with concentrations of 0.1-100 μM) is the logarithm of the dispersed (monomeric) protein volume fraction, $\mu = \ln(\phi_M)$, as seen in Supplementary Note S1, of the SI Sec. I. 2. The solubility energy, $\epsilon_B$, represents the combined effects of two factors: (a) the favorable interaction energy between amino acids in the dense cluster core (per protein) and (b) the entropic cost associated with the reduced conformational space of the protein in the 'hydrophobic' state compared to the 'water-soluble' configurations. Since cluster formation is favorable in the salinity and temperature regime studied, the solubility energy is negative ($\epsilon_B < 0$).

The core energy is a function of the number of proteins in the cluster $N$ while the shell energy is a function of its radius $R$. To present the cluster formation energy as a function of one geometrical parameter, we link these two with

$$N = \frac{4\pi}{3}\frac{R^3}{v}. \qquad (3)$$

The protein volume, $v$, is assumed to be the same in the core, shell, and solution. This simplification avoids introducing additional fitting parameters. With this simplification, the relation between protein volume fraction, $\phi$, to the protein concentration, $C$, is given by

$$\phi = v \cdot C. \tag{4}$$

For the subsequent calculations, we estimate $v$ in all configurations by modeling the protein as an ideal polymer. The radius of gyration, $R_G$ is given by $R_G = l_{aa}(N_{aa}/6)^{1/2}$, where $N_{aa}$ is the number of amino acids and $l_{aa}$ is a typical length of each amino acid (0.36 nm). The protein volume is then expressed as $v = \frac{4\pi}{3} R_G^3$. However, it is important to note that this is a rough approximation. The actual $R_G$ in solution is likely larger, as the proteins in question have a large intrinsically disordered domain that typically follows the scaling law $R_G \sim N^{0.57}$ [30]. In contrast, proteins in the core are "collapsed" (in the sense of polymers) and occupy a smaller volume, in the extreme case $v \sim N$ (instead of $v \sim N^{\frac{3}{2}}$ in the solution). The proteins in the shell are more extended and occupy an intermediate volume, situated between the extremes of the core and soluble configurations.

An advantage of our mesoscale approach is that these details do not affect the scaling laws that relate the most likely cluster size to the total protein concentration. However, concentration gradients (for example, between the core and shell) will introduce corrections to the cluster formation energy. A comprehensive analysis of these corrections lies beyond the scope of the present work.

The energy of forming a cluster of radii $R$ is given by the sum of Eqs. 1 and 2

$$U(R) = U_{core} + U_{shell} = \frac{4\pi}{3v}(\epsilon_B - \mu)R^3 + 4\pi\gamma R^2 - 8\pi\kappa_B J_s R + 8\pi\bar{\kappa}. \tag{5}$$

The balance between the minimum of $U(R)$, which favors a specific radius, and entropy that favors monomers as they maximize the translational entropy, gives the cluster number concentration (made dimensionless by multiplication by the protein molecular volume $v$) [22]

$$P(R) = \frac{v \cdot n_R}{V} = \exp\left[-\frac{U(R)}{k_B T}\right]. \tag{6}$$

$V$ is the system volume and $n_R$ is the number of clusters of size $R$. We note that this theory does not apply to clusters whose radius is comparable to the monomer size, as our mesoscopic approach used here fails in that regime. In fact, no clusters with a radius smaller than 15 nm are observed in all protein types and experimental methods discussed here (typical protein radius is 2.5-5 nm), indicating that the formation of clusters smaller than a critical radius is energetically unfavorable. This is possibly due to the high bending energy cost per cluster. This observation highlights the contrast to classical self-assembly (SN Sec. I 5).

We use this theoretical model to derive an analytical description of the size distribution of clusters and the mean size, $R^*$, as a function of the core and surface properties, and the total protein concentration in the system. This theory is compared to the experimental observations.

**Results**

*Cluster mean radius as a function of protein concentration*

We first address the most common cluster radius, $R^*$, of CPEB4 and FUS clusters as a function of total protein concentration, $C$. These are found based on the peaks of size distribution derived from Dynamic Light Scattering (DLS) for CPEB4, which is presented in Oranges *et al.* [11], and FUS clusters Nanoparticle Tracking Analysis (NTA) presented in Kar *et al.* [4]. Based on these experimental results, FUS and CPEB4 obey two different paradigms: $R^*$ of FUS clusters increases significantly with $C$ (Fig. 2A, open circles), and most (~99%) of the proteins are dispersed in solution as monomers [4]. In contrast, $R^*$ of CPEB4 clusters is almost unchanged as $C$ increases (Fig. 2B, open circles), and the clusters contain about 90% of the total protein concentration [11].

These behaviors agree with those of self-assembly above the CAC for CPEB4 and below the CAC for FUS: below the CAC, the fraction of protein in clusters is small, and the most common cluster radius, $R^*$, varies strongly as a power law of the protein concentration, $C$. Above the CAC, most of the proteins are found in clusters and $R^*$ varies only slightly with $C$.

We use self-assembly theory (Eqs. 1-3 and SN) to predict the properties of the cluster distribution both below and above the CAC and compare them with the experiments. From Eq. 6 (SN section I.1), we write the cluster number concentration in terms of the cluster aggregation number, $N$,

$$P(N) = \exp[-N\epsilon_N + N \ln(\phi_m)]. \tag{7}$$

$\phi_m$ is the protein monomer volume fraction, and $\epsilon_N$ is the energy per protein monomer in a cluster of size $N$. This can be related to the parameters introduced in Eqs. 1 and 2 by considering the total number of proteins in a cluster (Eq. 3),

$$\epsilon_N = \epsilon_B + 4\pi\gamma \left(\frac{3v}{4\pi}\right)^{\frac{2}{3}} N^{-\frac{1}{3}} - 8\pi\kappa J_s \left(\frac{3v}{4\pi}\right)^{\frac{1}{3}} N^{-\frac{2}{3}} + 8\pi\bar{\kappa} \cdot N^{-1}. \qquad (8)$$

The conservation of protein number in the system constrains $\phi_m$ and the total protein volume fraction $\phi$ via the volume fraction of proteins in clusters $\phi_c$:

$$\text{(a)} \ \phi = \phi_m + \phi_c, \quad \text{(b)} \ \phi_c = \sum_{N=2}^{N=\infty} N \cdot P(N) \qquad (9)$$

Eq. 7 can then be written as a self-consistent equation for the cluster number distribution (SN section I.1),

$$P(N) = \exp[-N\epsilon_N + N \ln \phi] \cdot \left(1 - \sum_{j=2}^{j=\infty} \frac{j \cdot P(j)}{\phi}\right)^N. \qquad (10)$$

We solve Eq. 10 for the most probable value of $N = N^*$ using a saddle point approximation of the cluster energy, which predicts the number of proteins in the most probable clusters (see SN section I.3). This approximation is valid as long as the distribution width is relatively narrow.

In addition, from our fits below, we estimate that the surface contributions of the tension and spontaneous curvature are much smaller than the bending contributions, $8\pi\bar{\kappa}$. Thus, in estimating $N^*$, we neglect the first two terms in Eq. 1 in a zeroth-order approximation. We note that this applies to the estimation of $N^*$ while the fits of the entire normalized distribution shown in the next section are insensitive to the bending contributions.

Taking the limit of large clusters ($N \gg 1$), we find that the concentration dependence of $N^*(\phi)$ below the CAC varies linearly with the protein volume fraction $\phi$ (see Eq. S45 in SN):

$$\ln N^* = \frac{8\pi\bar{\kappa}}{k_B T} + 1 + \log \phi. \qquad (11)$$

With Eqs. 3 and 4, Eq. 11 can be used to fit the theory to experimental data

$$R^* = \left(\frac{3}{4\pi}\right)^{\frac{1}{3}} \exp\left[\frac{8\pi\bar{\kappa}}{3k_BT} + \frac{1}{3}\right] v^{\frac{2}{3}} \cdot C^{\frac{1}{3}}. \tag{12}$$

In Fig. 2A, we show $R^*$, which was obtained from FUS NTA measurements (Fig. 4A of Kar, M. et al. (2022)) versus the FUS concentration $C$ (Fig. 2A, open circles). The fit (with *one* fitting parameter) agrees well with the theory. The inset shows the consistency with the power law $N^* \sim R^{*3} \sim C \sim \phi$ (Eq. 11). We note that the largest deviation from the theoretical fit is at 0.125 µM, where the distribution of cluster size is maximal, and our approximations are not accurate. From the fitting parameter (the slope in Eq. 12, $\left(\frac{3}{4\pi}\right)^{\frac{1}{3}} \exp\left[\frac{8\pi\bar{\kappa}}{3k_BT} + \frac{1}{3}\right] v^{\frac{2}{3}}$), we found that $8\pi\bar{\kappa}$ is roughly 16 $k_B$T for FUS.

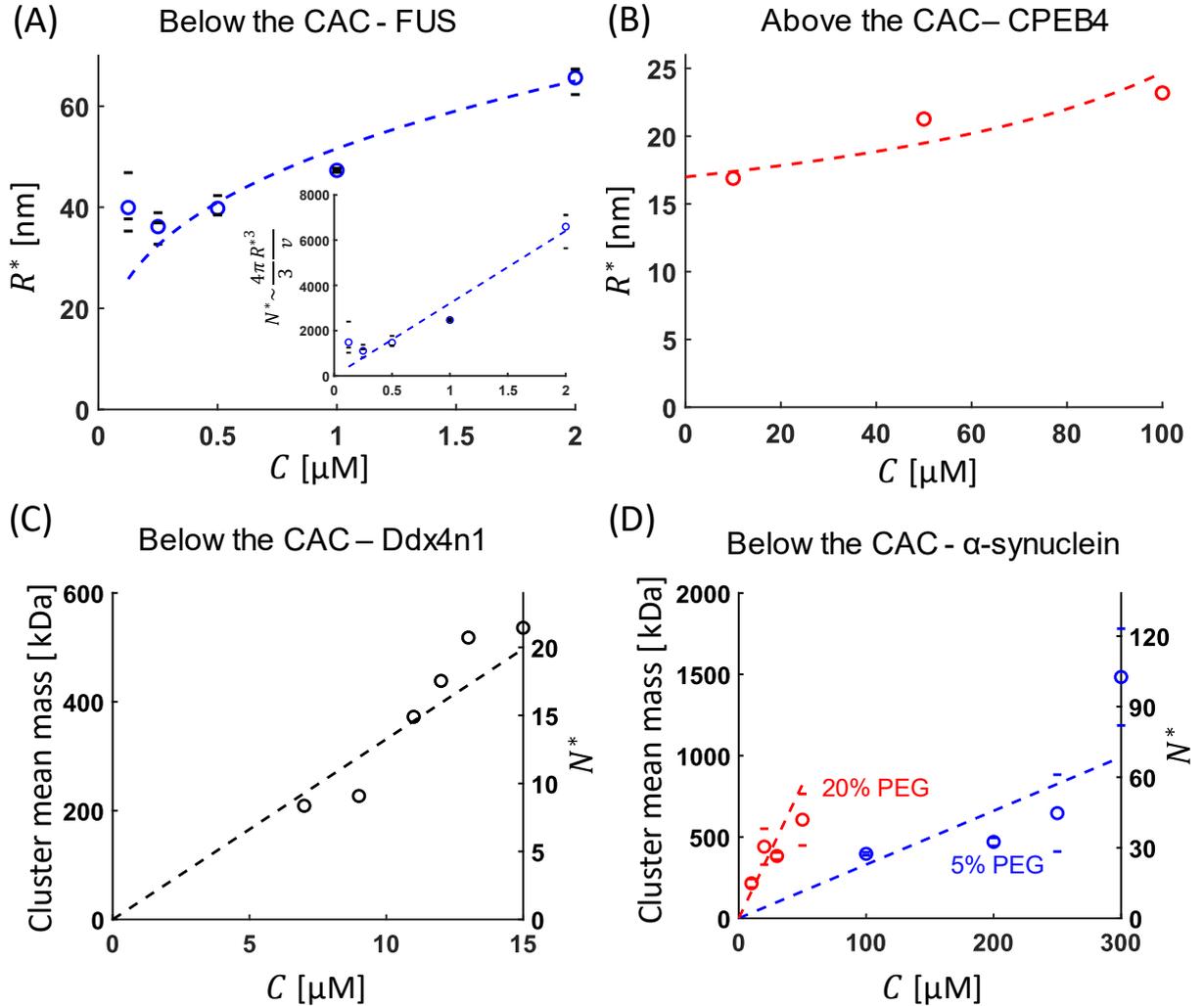

**Figure 2: Scaling law for the most common cluster size.** The most common cluster size plotted as a function of total protein concentration, $C$. Theoretical fits are represented by dashed lines, and experimental data by circles. **(A)** FUS cluster radius obtained from the NTA measurements, which are most sensitive to the smaller clusters. Theoretical fit to $R^* = a \cdot C^{\frac{1}{3}}$ (corresponding to Eq. 12), the fitting

parameter $a$ is found to be 5.25·10³ nm². Data for $C$ and $R^*$ are as listed in Table S3. The inset is the same data however the y-axis was rescaled to show the linear relation between $N^*$ (estimated as $\frac{4\pi}{3}\frac{R^{*3}}{v}$, with $v$ the estimated volume of FUS monomer, taken as 180 nm³) and $C$ as indicated in Eq. 13. The solid black lines are the $R^*$ (or $N^*$ in the insert) values taken from three different measurements, these are not error bars. The mean value of these is the blue circles **(B)** CPEB4: Data for $C$ and $R^*$ are as listed in Table S2. The theoretical curve is derived using a linear fit to $R^{*-3} = a + b \cdot C$, as described in Eq. 14. $a$ and $b$ were found to be 2·10⁻⁴ nm⁻³ and -0.018 nm⁻³µM⁻¹. The curve is generated by inverting the relationship, $R^* = (a + b \cdot C)^{-\frac{1}{3}}$ **(C and D)** Ddx4n1 and α-synuclein: cluster mass data taken from Ray et al (2023). This measurement is also most accurate for the smaller clusters (in the range of 15 kDa to 5 MDa, corresponding to $N$ in the range of 10⁰-10²). $N^*$ is the number of monomers in a cluster calculated as $N^* = M_{app}/M_{mono}$. The fitting follows a linear relationship, $N^* = a \cdot v \cdot C$ (Eq. 13), where $a$ is the fitting parameter and $v$ is the monomer volume (48 nm³ for Ddx4n1 and 22 nm³ for α-synuclein). The values of $a$ were found to be 4.63·10⁴ for Ddx4n1, 1.72·10⁴ for α-synuclein at 20% PEG, and 8.59·10⁴ at 5% PEG. **(D)** Two sets of measurements were done for α-synuclein, represented by the solid lines (that are not error bars). The average is the open circles.

In addition to FUS, α-synuclein, and Ddx4n1 proteins were recently reported to form clusters whose size strongly depends on the protein concentration and contain only a small fraction of the total protein in the system [13]. This behavior is indicative of a system below the CAC and aligns with the scaling relationship presented in Eq. 11, $N^* \sim \phi$. We fitted the mass of these clusters (measured using mass photometry [13]), proportional to $N^*$, to the linear relation obtained from Eq. 11

$$N^* = \frac{M_{cluster}}{M_{mono}} = \exp\left[\frac{8\pi\bar{\kappa}}{k_BT} + 1\right] \cdot v \cdot C. \qquad (13)$$

Here $M_{mono}$ is the monomer mass, 14.46 kDa for α-synuclein and 25 kDa for Ddx4n1. The results are presented in Fig. 2C for Ddx4n1 and 2D for α-synuclein at 5% and 20% PEG concentrations and show a good agreement with the model. From the slope ($\exp\left[\frac{8\pi\bar{\kappa}}{k_BT} + 1\right]$, 4.63·10⁴ for Ddx4n1, 1.72·10⁴ for α-synuclein at 20% PEG, and 8.59·10⁴ at 5% PEG) and the estimated volume of the protein monomers (48 nm³ for Ddx4n1 and 22 nm³ for α-synuclein), we find that the contribution of bending rigidity to the formation of these clusters ($8\pi\bar{\kappa}$) is 9.7 k_BT for Ddx4n1, 10.4 k_BT for α-synuclein with 20% PEG, and 8.8 k_BT at 5% PEG. These are comparable to the bending rigidity energy of FUS.

We note that the number of monomers in the Ddx4n1 and α-synuclein clusters (a few dozen, Fig. 2C and 2D) is significantly lower than those in FUS or CPEB4 clusters ($10^2$–$10^3$ monomers). This difference makes the continuum core-shell model presented here less applicable and could explain the observed deviation from linear behavior for small clusters. Nevertheless, our results indicate that these clusters behave as swollen micelles (microemulsions) below the CAC with a significant bending rigidity, supporting the suggestion by Ray et al. (2023) that they exhibit a micelle-like structure [13].

In contrast to FUS, Ddx4n1 α-synuclein and other proteins of the FET family (presented in the SI section III: Supplementary Note S2, Supplementary Fig. S4 and Supplementary Figs. S5-S7), the value of $R^*$ for CPEB4 clusters increases slowly with protein concentration, indicative of a system above the CAC where most of the added protein forms additional clusters. In Supplementary Note S1, of section I. 6 of the SI), we show that under such conditions and near the CAC, the most probable cluster radius varies with the protein concentration as (Eq. S65):

$$\frac{R_{CAC}^3}{R^{*3}} = 1 + \frac{1}{f_{CAC}}\left(1 - \frac{C}{C_{CAC}}\right). \tag{14}$$

With $C_{CAC}$, $R_{CAC}$ and $f_{CAC}$ the concentration (usually given in Molar in experiments), the most probable cluster size, and the protein fraction in clusters at the CAC, respectively.

We fit this theoretical prediction to the size of CEPB4 clusters as a function of total protein concentration in Fig. 2B and found good agreement. Unfortunately, there are only three data points for CPEB4, so the details of the fit are not statistically significant. Nevertheless, we estimate the bending rigidity of these clusters by rearranging Eq. S55 (noting that $\epsilon_T = 8\pi\bar{\kappa}$), correlating the concentration and radius at the CAC and the bending rigidity energy,

$$8\pi\bar{\kappa} = \ln\frac{4\pi R_{CAC}^3}{3v^2 \cdot f_{CAC} \cdot C_{CAC}} - 1. \tag{15}$$

We do not know the concentration, cluster radii, or their volume fractions at the CAC, but we can estimate them by considering the values measured at 10 µM (which is the closest measurement to CAC reported). With these values and assuming $f_{CAC} = 0.9$ (90% of the protein was estimated to be in clusters [11]), a cluster radius of 17 nm (Table S2), and a CPEB4 monomer volume of 125 nm$^3$, the bending rigidity is ($8\pi\bar{\kappa}$) 11 k$_B$T. To compare, the bending energy per cluster of FUS was estimated above to be 16 k$_B$T, a significant 5 k$_B$T difference since it enters into the exponential that characterizes the distribution (Eq. 12).

We estimate the bending modulus of the different clusters by considering $\kappa_B \sim -\kappa_G$ as in typical amphiphilic layers [21]. With that, we find the bending modulus, $\kappa_B$, of FUS as 1.3 k$_B$T, CPEB4 is 0.9 k$_B$T, Ddx4n1 is 0.8 k$_B$T, α-synuclein at 20% PEG is 0.8 k$_B$T and 0.7 k$_B$T at 5% PEG. This is consistent with an experimental estimate for micron-scale LLPS domains in stress granules [31]. We do not expect $\kappa_B$ to be different for mesoscale or micron-scale clusters since it is a material property of the interface and is independent of the shape [21].

Despite the approximate nature of these fits, the qualitative difference between CPEB4 and FUS (and the others) for the concentration dependence of the most probable aggregation number $N^*$ or radius $R^*$ is clear; CPEB4 shows a very weak dependence, characteristic of a self-assembling system above its CAC, while FUS and the others (measured by NTA or cluster mass photometry, both sensitive to the smaller cluster – in contrast to the FUS DLS data), show a strong dependence, indicative of systems below their CAC. This is striking and consistent with the experimental observation that most CPEB4 proteins are found in clusters (above the CAC), while most FUS proteins are dispersed in solution (below the CAC). These differences are primarily due to the 5 k$_B$T difference in their bending energies per cluster; the larger energy for FUS reduces the probability of the most probable cluster by a factor of $e^{-5} \sim 0.007$ compared with CPEB4. This assumes that the contributions of the other energies per cluster are small compared with the bending energy contributions, which for both proteins are greater than 10 k$_B$T. This indeed is what we find in the following sections.

*Analysis of the cluster size distribution to infer core and surface energies*

Next, we analyze the cluster size distribution using our core-shell model (Eqs. 1-3) to infer the core energy, surface tension, and spontaneous curvature of the clusters. To eliminate terms that are independent of the cluster size, we normalize the cluster number distribution to its peak value at $P(R = R^*)$ that was derived in the previous section. With this normalization, the number distribution can be written as a function of only three parameters (full derivation in Supplementary Information of Oranges et al. [11], where this approach was used to analyze temperature dependence of size distribution)

$$\frac{P(R)}{P(R^*)} = \exp\left[-A_3\left(\frac{R^3}{R^{*3}} - 1\right) - A_2\left(\frac{R^2}{R^{*2}} - 1\right) - A_1\left(\frac{R}{R^*} - 1\right)\right] \quad (16)$$

With

$$\text{(a) } A_1 = -\frac{8\pi \kappa_B J_s R^*}{k_B T}, \quad \text{(b) } A_2 = \frac{4\pi \gamma R^{*2}}{k_B T}, \quad \text{(c) } A_3 = \frac{4\pi}{3} \frac{R^{*3}}{v} \frac{(\epsilon_B - \ln \phi_m)}{k_B T}. \tag{17}$$

$A_1$ is a measure of the spontaneous curvature, which we denote as 'curvature tendency', with positive values ($J_s < 0$) signifying a bulkier hydrophobic blob and a smaller hydrophilic blob, and the opposite for negative values of $A_1$ ($J_s > 0$) [22, 32]. $A_2$ is a measure of the residual interfacial tension energy that arises from less-closely packed regions in the 'amphiphilic' layer where there is hydrophobic (core) – hydrophilic (solvent) contact, and $A_3$ is a measure of the core energy, all normalized by the thermal energy k$_B$T. The bending rigidity $\bar{\kappa}$ calculated in the previous section cannot be inferred from these fits, since its contribution to the energy per cluster is independent of $R$ and is eliminated by the normalization.

We find $A_1, A_2,$ and $A_3$ by performing a least squares minimization of $\ln\left[\frac{P(R)}{P(R^*)}\right]$ since its linearity with the parameters simplifies the procedure:

$$\chi^2_{min} = \min \sum_{i=1}^{i=M} \left[\ln\left(\frac{P(R_i)}{P(R^*)}\right) + A_3\left(\frac{R_i^3}{R^{*3}} - 1\right) + A_2\left(\frac{R_i^2}{R^{*2}} - 1\right) + A_1\left(\frac{R_i}{R^*} - 1\right)\right]^2. \tag{18}$$

Here, $M$ is the number of DLS data points, $R_i$ is the cluster radius, and $P(R_i)$ is the cluster number concentration of the radius $R_i$.

We fit the experimental data to the theoretical model exclusively in the region where the normalized cluster size distribution satisfies $R > R^*$. This regime was chosen because larger clusters exhibit greater scattering cross-sections, resulting in a more accurate measurement in the DLS data [33]. Additionally, our mesoscopic theory is more applicable to the regime of larger radii, since we use continuum concepts such as interfacial tension and curvature energy, which assume that the cluster radius is much larger than the molecular size. Specifically, CPEB4 proteins have a typical diameter of 3–5 nm, while the cluster radius $R^*$ is approximately 20 nm (Table S2). Consequently, clusters with $R < R^*$ have a radius of curvature comparable to the protein diameter and are not adequately described by our model. The limitations of fitting at $R < R^*$ range and a representative example that includes the fit for smaller cluster radii is discussed in more detail in Supplementary Note S3 and Supplementary Figure S5 in Sec. IV of the SI.

The parameter values obtained from the fits presented in Figure 3 (and complementary Figure S2) are found in Tables S2 and S3 for the DLS number distribution of CPEB4 and FUS clusters (sensitive to the larger clusters) and Table S4 for the NTA measurements in FUS (sensitive to the smaller clusters). Due to the small number of data points and the experimental noise, the fitting results reported here should be considered indicative only of the trend with increasing protein concentration.

Analysis of the clusters' size distributions of other proteins of the FET family (EWSR1, TAF15, and hnRNPA3), which are also well-described by our model, are presented in Supplementary Note S2, Supplementary Figure S4, Supplementary Figs. S5-S7 of Sec. III of the SI.

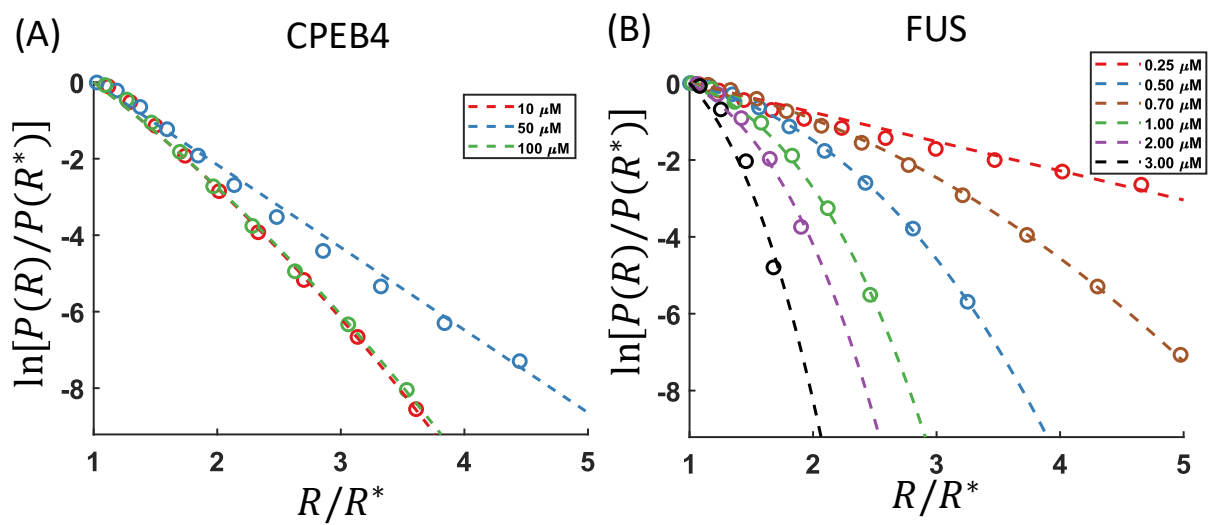

**Figure 3 – Normalized Dynamic Light Scattering (DLS) number distribution**. Open circles – experimental data, dashed lines – theoretical fit to Eq. 16. The protein concentration (in Molar) is indicated in the legends. **(A)** CPEB4 at 277 K and 100 mM NaCl. Data taken from Oranges *et al.* (2024). The fitting result is summarized in Table S2, and a linear-linear plot is presented in Fig. S2A. **(B)** FUS at 298 °K, 20 mM Tris, pH 7.4, and 100 mM KCl. Data taken from Kar et al. (2022). The result of the fitting is summarized in Table S3. A linear-linear plot is presented in Fig. S2 B.

Except for points $R \approx R^*$, the CPEB4 data for the log of the normalized distribution shown in Fig. 3A is quite linear in the cluster radius, which signifies relatively small core ($A_3$) and interfacial tension ($A_2$) energies compared with the curvature tendency contribution ($A_1$) whose contribution to $\ln(P(R)/P(R^*))$ is linear in $R$. A similar trend is observed in the temperature dependence of CPEB4, as shown in Fig. 9C of Oranges et al. (2024). In other words, the CPEB4 cluster size distribution is dominated by the curvature tendency of the proteins in the shell. Large clusters have a decreased probability due to their actual curvature, whose sign is opposite to the preferred curvature of the shell proteins ($A_1 > 0$).

The numerical fits support this qualitative observation: CPEB4 clusters have negligible core energy (Table S2). The interfacial tension energy is small (~0.3 $k_B T$ corresponding to interfacial tension of 0.2-0.3 µN/m), and the curvature tendency term is the dominant term, fitted to 1.9-2.1 $k_B T$. The cluster properties are almost unchanged even as the concentration is increased by an order of magnitude (from 10 µM to 100 µM), consistent with a system above the CAC.

Using the estimated bending rigidity of CPEB4 clusters from the previous section (1.1 $k_B T$) and the fitted values of $A_1$ (Table S2), we estimate the spontaneous curvature, $J_s$ (Eq. 17 A), of CPEB4 proteins on the shell in the range of -4 µm$^{-1}$ to -3 µm$^{-1}$. This is about an order of magnitude smaller than a previous estimate based solely on the relative sizes of the hydrophobic and hydrophilic blobs of CPEB4 [11], indicating that the blobs are highly deformable and soft.

In contrast to CPEB4, the DLS data for FUS suggest that $\ln(P(R)/P(R^*))$ of large clusters at high protein concentration is dominated by parabolic and cubic terms in the cluster radius, while $\ln(P(R)/P(R^*))$ for the small clusters measured by NTA are linear, as CPEB4. This indicates that the larger clusters have non-vanishing interfacial tension ($A_2$) and\or core energy ($A_3$), while the smaller ones are dominated by their curvature tendency ($A_1$). This suggests that the large clusters, particularly those close to LLPS (occurring at 3 µM [4]), are fluctuations and rather unstable, characteristic of a system below the CAC, and consistent with the experiments and theory for the concentration dependence of $R^*$ in the previous section.

We found, based on our fits to FUS DLS data (Fig. 3 B and Table S3), that $A_1$ decreases from 0.76 $k_B T$ (1/3 of CPEB4) at 0.25 µM to vanishing values at 0.7 µM. This is reasonable since one cannot extract from the distribution of such large clusters any tendency to bend on significantly smaller scales. The core and surface tension energies, which are hard to numerically distinguish due to the limited number of available data points (especially for the data sets at 2 µM and 3 µM), increase from vanishing values at 0.25 µM to 1.2 $k_B T$ at 3 µM. The tension and core energies dominate the distribution in large clusters as they scale as $R^2$ and $R^3$, respectively ($R^*$ is 49 nm at 0.25 µM and 381 nm at 3 µM). The fits to NTA measurements of FUS (Fig. 4 and Table S4), which are most sensitive to the radii of small clusters, indicate that the core and tension energies are approximately zero. The small FUS clusters are dominated by curvature tendency (2 $k_BT$ at 0.25 µM decreasing to 0.8 $k_BT$ at 2 µM), similar to the CPEB4 clusters.

Based on these results, we estimate the FUS clusters' interfacial tension is in the $10^{-3}$-$10^{-2}$ µN/m range (Tables S3 and S4, for large and small clusters, respectively), indicating a tight packing of the 'amphiphiles' configuration on the clusters' surface. These tension values are an order of magnitude smaller than those of CPEB4 (~$10^{-1}$ µN/m, Table S2) and are the likely reason that allows the cluster size to increase as proteins are added to the system, since increasing the surface area involves only a negligible free energy cost. Based on the bending rigidity of FUS clusters estimated in the previous section (1.3 $k_BT$) and the fits to the NTA data (Table S4) that represent the curvature tendency of the clusters more accurately than DLS, we estimate the spontaneous curvature, $J_s$, of the FUS proteins to be between -1.7 µm$^{-1}$ to -2.3 µm$^{-1}$, similar in magnitude to that of CPEB4.

Finally, we calculate the solubility energy, $\epsilon_B$, which is the energy difference between a protein molecule in the core of the cluster relative to the aqueous solution, based on the fitted parameter $A_3$ (Tables S3 and S4) and the monomer volume fraction, $\phi_m$. Kar *et al.* estimated that 0.15% of the proteins are in clusters at 0.25 µM [4]. The fit of our model to the DLS and NTA data showed that at this protein concentration $A_3 \ll k_BT$, meaning that $\epsilon_B \cong \ln \phi_m$ (Eq. 17 C). The protein volume fraction, $\phi_m$, is estimated to be 2.7·10$^{-5}$, and the solubility energy we estimate is thus -10.5 $k_BT$. To compare, the CPEB4 solubility energy was estimated as -7.5 $k_BT$ [11]. We note that this is an energy per protein molecule and thus coarse-grains over all the amino acids; converting it to an energy per amino acid gives only a fraction of $k_BT$.

To conclude, the normalized size distribution of the FUS clusters changes strongly with increasing protein concentration, shifting the distribution from being dominated by the curvature tendency at low concentrations (small clusters) to being dominated by core and\or tension energies at high concentrations (large clusters). CPEB4 clusters are almost unchanged even as the protein concentration increases by an order of magnitude and are always dominated by their curvature tendencies. The behavior of FUS is similar to that of a microemulsion below the CAC, while that of CPEB4 is similar to a swollen micelle (microemulsion) above the CAC.

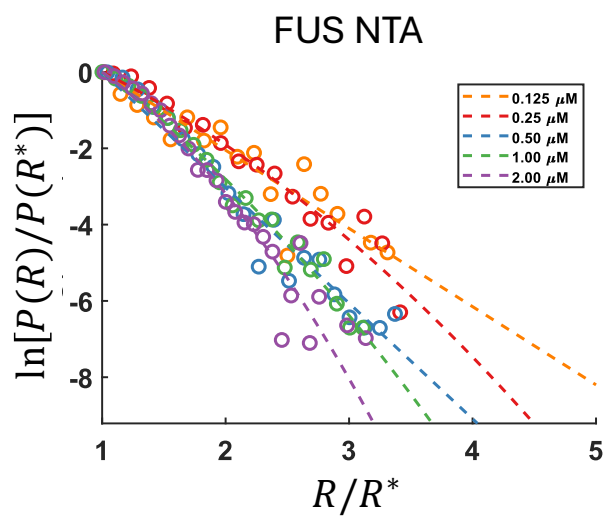

**Figure 4 – Normalized FUS Nanoparticle Tracking Analysis (NTA).** Data taken from Kar et al.[4]. Open circles – experimental data, dashed lines – theoretical fit to Eq. 16. The protein concentration is indicated in the legends. The transition to LLPS occurs at a concentration of approximately 3 μM. The result of the fitting is summarized in Table S4. A linear-linear plot is presented in Fig. S2 C. Measurement was done at 298 °K, 20 mM Tris, pH 7.4, and 100 mM KCl.

## Discussion

The analysis of the measurements of the most common cluster size $R^*$ as a function of protein concentration as well as the fits to the cluster size distributions indicate that FUS and CPEB4 represent two different paradigms for cluster formation in the one-phase regime: below and above the CAC, respectively. This qualitative difference is due to the difference in surface properties of these clusters, which we inferred from various experiments in this work based on the theoretical model we presented here.

The protein fraction in clusters is mostly determined by the solubility energy (Eq. 2), while the cluster size distribution is determined by the surface (shell) energy (Eq. 1). The $R^*$ size scaling obtained from FUS NTA, CPEB4 DLS, and mass photometry of Ddx4n1, and α-synuclein measurement (Fig. 2) showed that the bending rigidity energy is 10-16 $k_BT$ per cluster. These measurements (in contrast to the FUS DLS) are appropriate for relatively small clusters. The bending rigidity modulus estimated here is of the order of 1 $k_BT$ for all these types of clusters, which is comparable to bending rigidities measured for other condensates [31]. The size distribution analysis around $R^*$ for FUS and CPEB4 clusters showed that all the other terms ($A_1$, $A_2$ and $A_3$, Eq. 17) are order 1-2 $k_BT$ far from LLPS (Table S2-S4). This means that the surface energy of small clusters is dominated by their bending rigidities. The ultra-low tensions we report here (Tables S2-S7) are in agreement with measurements on condensates [31, 34, 35].

We note that the scaling of $R^*$ presented in Eqs. 11-15 holds only if the energy per cluster is constant or weakly changing. The large FUS clusters measured using DLS (Table S3), particularly close to LLPS, do not meet this condition as the tension and\or core energies significantly change with cluster size. Therefore, we estimated the bending rigidity using NTA measurement of FUS clusters (Fig. 2), which exclude the very large clusters that do not fill this requirement. Our estimation of bending rigidity modulus is also valid for large clusters since the bending energy is independent of the size.

The formation energy (sum of all the energies of the model given by Eq. 5) of small, FUS clusters (far from LLPS) is 17 $k_BT$, while CPEB4 clusters are less costly and are roughly 13 $k_BT$. This energy difference of approximately 4 $k_BT$ per cluster is mostly due to the bending energy ($8\pi\bar{\kappa}$) and accounts for the relative fraction of proteins in clusters, with a ratio of ~90 (~1% for FUS and ~90% for CPEB4), of the same order as the relative Boltzmann factors of $e^4 \approx 55$. In contrast to the total formation energy (including also the surface), FUS clusters' solubility energy per protein molecule, $\epsilon_B$, is 3 $k_BT$ lower compared to CPEB4.

The higher bending energy of FUS clusters (16 $k_BT$ compared to 11 $k_BT$ in CPEB4) indicates a tighter packing and more aligned ordering of the proteins in the 'amphiphilic configuration' (Fig. 1), reducing the interfacial tension from the exposure of hydrophobic amino acids in the core to the water. The size distribution analysis around $R^*$ (Figs. 3 and 4) obtained by DLS and NTA measurements supports this prediction: These showed that FUS clusters' surface tension (Tables S3 and S4) is an order of magnitude lower than that of CPEB4 (Table S2). The ultra-low tension (~ 100 nN/m, 5 orders of magnitude smaller than the water-oil tension of ~10 mN/m) and relatively high bending rigidity energy of FUS clusters are consistent with a system below the CAC: clusters are rare due to their high formation energy, but grow with little free energy penalty, meaning they exist as fluctuations. In contrast, CPEB4 clusters have a lower bending rigidity and higher interfacial tensions, so these favor the formation of more clusters over the growth of existing ones. This is consistent with relatively stable clusters as in a system above its CAC.

Further, our analysis suggests that the energy of a molecule in a cluster relative to the aqueous solution is less than $k_BT$ per molecule (clusters contain $10^2$-$10^3$ molecules). This is what might be expected for a non-folded, disordered protein, as typical amino-acid level interactions are only several k$_B$T [36]. Of course, the stickers discussed in Kar et al. [4] most probably coarse-grained over several amino acids, resulting in an effectively large interaction. We emphasize that the goal of our model is not to determine the interactions at the amino-acid level, but rather to account for the mesoscale properties of the disordered protein as a whole in a coarse-grained model, and we find those energies to be of the order of k$_B$T for the entire cluster.

We note that the spontaneous curvature of both cluster types considered here is negative. That is, the proteins' preferred packing is opposite to the actual packing in the cluster shell. With the vanishingly low interfacial tension, this energy cost per cluster suppresses the formation of larger clusters in the distribution. It is tempting to speculate that the relative packing of the hydrophobic and hydrophilic blobs of certain intrinsically disordered proteins is the major factor determining the ability to form clusters that precede LLPS. While the proteins in the shell have an amphiphilic conformation, they must have a negative spontaneous curvature, which limits clusters larger than a certain size.

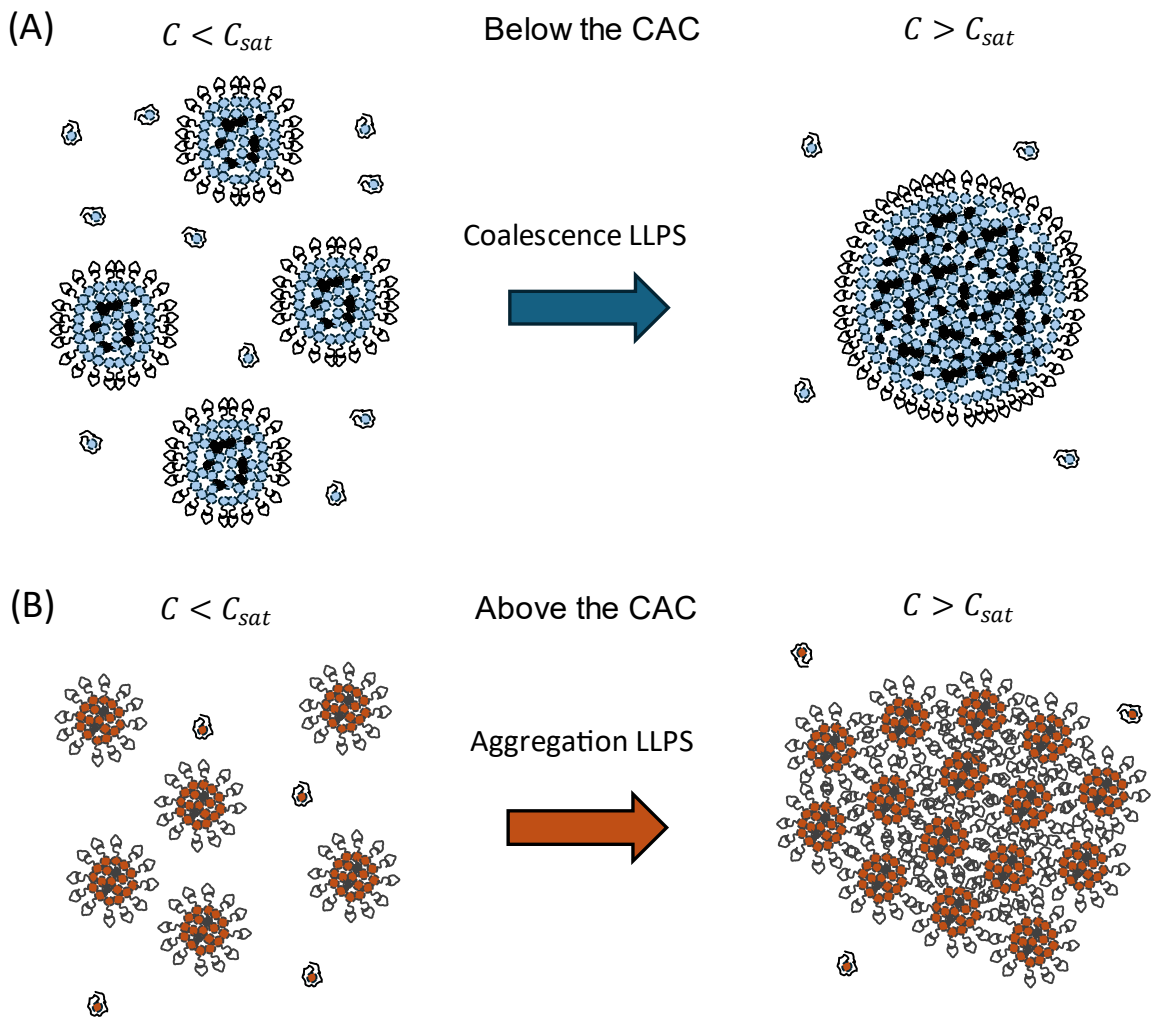

**Figure 5 – Liquid-liquid phase separation (LLPS) pathways. (A)** Coalescence LLPS: The merger of clusters into a micron-scale LLPS domain occurs via fusion. LLPS occurs at a critical cluster radius, where the interfacial tension energy exceeds the cluster's translational entropy. In addition, the monomers in solution must change from a hydrophilic to a hydrophobic dominant conformation. **(B)** Aggregation LLPS: The formation of a large cluster is mediated by an attractive interaction between the intrinsically disordered proteins of the proteins at the amphiphilic configuration. Here, the transition occurs at a critical cluster concentration and depends strongly on the length and interaction energy of the intrinsically disordered proteins.

We speculate that FUS and CPEB4 may follow distinct pathways to LLPS, characterized by different internal structures within the micron-scale LLPS domains (Fig. 5). FUS clusters grow in size with total protein concentration (Table S3). At the saturation concentration of 3 µM, there is a coexistence of micron-size (LLPS) domains, clusters, and a very large fraction of monomers (Fig. 8 of Kar et al. [4]). The LLPS domains grow via cluster coalescence and become larger as a function of time. This is consistent with the theoretical fits presented here to the DLS data (Table S3): These large clusters, which have costly core and/or surface energies, tend to undergo fusion of their cores to reduce those energies (Fig. 5 A). LLPS micron-sized domains consist of fused clusters, which essentially contain only the cluster core. This pathway to LLPS via cluster coalescence is also supported by recent work on a FUS-MBP system showing that FUS-MBP LLPS occurs through the coalescence of FUS clusters [12].

It is also possible, in principle, that in some systems, the pathway to LLPS is dominated by a drastic change in the dominant conformation of the protein monomers in solution from hydrophilic to hydrophobic without any significant intermediate clusters. Both pathways have the same final configuration – a large macroscopic LLPS domain with most proteins in hydrophobic configurations. A more comprehensive determination of the pathway to LLPS for a broader range of temperatures and concentrations would shed further light on the dominance of cluster coalescence versus monomer associations in FUS and related systems.

In contrast to FUS, CPEB4 clusters grow very slowly with increasing protein concentration, with negligible cluster tension energy even close to LLPS. This system is also significantly more concentrated, with the CPEB4 concentration in the cluster regime (that precedes LLPS) being two orders of magnitude larger than that of FUS and the other proteins in the FET family (100 µM for CPEB4 versus 3 µM for FUS). In addition, EPR experiments showed that the fractions of proteins distributed among the core, shell, and dispersed monomers do not change abruptly at LLPS [5, 11], suggesting that the clusters remain intact during the transition. We, therefore, propose that for the paradigm of CPEB4 (above their CAC), LLPS will occur via aggregation (in contrast with fusion) of existing clusters at a critical concentration $C_{sat}$ (Fig. 5 B). A similar process was proposed to occur when the concentration of CPEB4 was held constant, but the temperature changed [11]. However, this hypothesis remains to be experimentally validated through more in-depth structural analysis of macroscopic CPEB4 condensates.

It is also possible that the protein configuration at the core of the clusters we consider here is not the 'true' free energy minimum and that the cluster core 'age', as seen in macroscopic FUS LLPS domains [37]. Some evidence for this came from Cabau and colleagues [27] who reported a progressive increase in CPEB4 cluster size from 55 nm to 90 nm over 15 hours, accompanied by a rise in the polydispersity index. They proposed that these multimers evolve into mesoscopic condensates, comparable to those observed by optical microscopy but significantly smaller. However, due to the limitations in physically characterizing these structures, the authors classified them as distinct species. We believe that the identity and nature of these evolving species remain an open question and should be investigated in greater detail before applying our theoretical framework to them.

To conclude, we have presented a theoretical model to describe the size distribution of protein clusters that precede LLPS and used it to analyze new and existing experimental data *in-vitro*. We propose that a similar approach could be used to model the behavior of condensates *in-vivo* as well as other condensate geometries, such as recently observed fibrils in certain FUS clusters [28] and Ebola-induced condensates [38, 39]. Direct measurement and modeling of the protein at the amino-acid level, which requires other techniques, are certainly of interest to verify our mesoscale approach.

The main novelty of our proposed theory lies in the extraction of the ultra-low tension of the clusters from their size distribution, as well as the cluster bending rigidity and spontaneous curvature, all of which are attributed to the amphiphilic conformation of the protein at the interface of the cluster core and the solvent. These properties should be considered when analyzing and modeling membrane-less organelles' equilibrium shapes and kinetics, such as fusion, fission, and shaping processes.

**Acknowledgments**

SS holds the Fern and Manfred Steinfeld chair and is grateful for a grant from the Volkswagen Foundation 197/98. DG acknowledges a grant from the Israel Science Foundation 2253/18. AAH acknowledges support from the NOMIS foundation and SS and AAH acknowledge the support of a Volkswagen foundation Life award. This research was supported in part by the Helen and Martin Kimmel Institute for Magnetic Resonance Research and the historic generosity of the Perlman Family Foundation (DG, SS). We are grateful for the discussions with Alexey Bogdanov, Hagen Hofmann, and Rohit Pappu.

**Author Contributions**

GG - model, calculations, data fits, interpretation, writing; MS, MK, AH, DG – experiments, interpretation, writing; SS – model, interpretation, writing.

**Competing Interests**

AA Hyman is founder and SAB member of Dewpoint therapeutics. The other authors have no competing interests.

**Data Availability**

Raw data to reproduce figures, along with the MATLAB files used for data analysis, are deposited at https://github.com/GonenGolani/protein-clusters-data-2024.

**Figure 1 – Protein Conformations. (A)** Illustration of conformations of dilute protein monomers in the solvent and those packed in the suggested core-shell structure of the clusters. The blue color denotes more hydrophobic parts, but does not imply that these segments are folded. The black, hydrophilic domain of the proteins in the cluster core is more collapsed. **(B)** The interfacial layer consists of more hydrophilic protein segments that face the solvent and more hydrophobic segments that face the core.

**Figure 2: Scaling law for the most common cluster size.** The most common cluster size plotted as a function of total protein concentration, $C$. Theoretical fits are represented by dashed lines, and experimental data by circles. **(A)** FUS cluster radius obtained from the NTA measurements, which are most sensitive to the smaller clusters. Theoretical fit to $R^* = a \cdot C^{\frac{1}{3}}$ (corresponding to Eq. 12), the fitting parameter $a$ is found to be 5.25·10$^3$ nm$^2$. Data for $C$ and $R^*$ are as listed in Table S3. The inset is the same data however the y-axis was rescaled to show the linear relation between $N^*$ (estimated as $\frac{4\pi}{3}\frac{R^{*3}}{v}$, with $v$ the estimated volume of FUS monomer, taken as 180 nm$^3$) and $C$ as indicated in Eq. 13. The solid black lines are the $R^*$ (or $N^*$ in the insert) values taken from three different measurements, these are not error bars. The mean value of these is the blue circles **(B)** CPEB4*:* Data for $C$ and $R^*$ are as listed in Table S2. The theoretical curve is derived using a linear fit to $R^{*-3} = a + b \cdot C$, as described in Eq. 14. $a$ and $b$ were found to be $2 \cdot 10^{-4}$ nm$^{-3}$ and $-0.018$ nm$^{-3}$μM$^{-1}$. The curve is generated by inverting the relationship, $R^* = (a + b \cdot C)^{-\frac{1}{3}}$ **(C and D)** Ddx4n1 and α-synuclein: cluster mass data taken from Ray et al (2023). This measurement is also most accurate for the smaller clusters (in the range of 15 kDa to 5 MDa, corresponding to $N$ in the range of $10^0$-$10^2$). $N^*$ is the number of monomers in a cluster calculated as $N^* = M_{app}/M_{mono}$. The fitting follows a linear relationship, $N^* = a \cdot v \cdot C$ (Eq. 13), where $a$ is the fitting parameter and $v$ is the monomer volume (48 nm³ for Ddx4n1 and 22 nm³ for α-synuclein). The values of $a$ were found to be $4.63 \cdot 10^4$ for Ddx4n1, $1.72 \cdot 10^4$ for α-synuclein at 20% PEG, and $8.59 \cdot 10^4$ at 5% PEG. **(D)** Two sets of measurements were done for α-synuclein, represented by the solid lines (that are not error bars). The average is the open circles.

**Figure 3 – Normalized Dynamic Light Scattering (DLS) number distribution**. Open circles – experimental data, dashed lines – theoretical fit to Eq. 16. The protein concentration (in Molar) is indicated in the legends. **(A)** CPEB4 at 277 K and 100 mM NaCl. Data taken from Oranges *et al.* (2024). The fitting result is summarized in Table S2, and a linear-linear plot is presented in Fig. S2A. **(B)** FUS at 298 °K, 20 mM Tris, pH 7.4, and 100 mM KCl. Data taken from Kat et al. (2022). The result of the fitting is summarized in Table S3. A linear-linear plot is presented in Fig. S2 B.

**Figure 4 – Normalized FUS Nanoparticle Tracking Analysis (NTA).** Data taken from Kar et al.[4]. Open circles – experimental data, dashed lines – theoretical fit to Eq. 16. The protein concentration is indicated in the legends. The transition to LLPS occurs at a concentration of approximately 3 μM. The result of the fitting is summarized in Table S4. A linear-linear plot is presented in Fig. S2C. Measurement was done at 298 °K, 20 mM Tris, pH 7.4, and 100 mM KCl.

**Figure 5 – Liquid-liquid phase separation (LLPS) pathways. (A)** Coalescence LLPS: The merger of clusters into a micron-scale LLPS domain occurs via fusion. LLPS occurs at a critical cluster radius, where the interfacial tension energy exceeds the cluster's translational entropy. In addition, the monomers in solution must change from a hydrophilic to a hydrophobic dominant conformation. **(B)** Aggregation LLPS: The formation of a large cluster is mediated by an attractive interaction between the intrinsically disordered proteins of the proteins at the amphiphilic configuration. Here, the transition occurs at a critical cluster concentration and depends strongly on the length and interaction energy of the intrinsically disordered proteins.

**Supplementary information:**

# Mesoscale properties of protein clusters determine the size and nature of liquid-liquid phase separation (LLPS)

Gonen Golani[1,2], Manas Seal[2], Mrityunjoy Kar[3,4], Anthony A. Hyman[3], Daniella Goldfarb[2] and Samuel Safran[2]

[1] Department of Physics, Haifa University, Haifa, Israel

[2] Department of Chemical and Biological Physics, Weizmann Institute of Science, Rehovot, Israel

[3] Max Planck Institute of Cell Biology and Genetics, Dresden, Germany

[4] Institute of Biofunctional Polymer Materials, Leibniz Institute of Polymer Research Dresden, Dresden, Germany

**Table of Contents:**

# I. Supplementary Note S1: General Theory for Cluster Size Distribution
## 1. Distribution of cluster sizes – General Formulation

We consider the total number of proteins, $N_{total}$, to be distributed in clusters:

$$N_{total} = n_1 + \sum_{N=2}^{N=\infty} n_N \cdot N. \tag{S1}$$

With $n_1$, the number of monomers and $n_N$ is the number of clusters of size $N$. The total protein volume fraction, $\phi$, in the system is related to the volume fractions of the monomers and clusters by:

$$\phi = \frac{N_{total}}{M} = \phi_m + \sum_{N=2}^{N=\infty} \frac{n_N \cdot N}{M}. \tag{S2}$$

$M = V/v$ is the ratio of the system volume, $V$, to the protein volume, $v$. We define the *number concentration* of clusters of size $N$ to be:

$$P_N = \frac{n_N}{M}. \tag{S3}$$

Thus, the total protein volume fraction is given by the sum of the monomers and clusters volume fractions ($\phi_m$ and $\phi_C$, respectively), Eq. S2:

$$\phi = \phi_m + \phi_C = \phi_m + \sum_{N=2}^{N=\infty} N \cdot P_N. \tag{S4}$$

We denote $E_N$ as the thermodynamic work of creating a cluster from $N$ non-interacting monomers in the solution. This energy does not include translational entropy, which we consider separately in the following. The energy per monomer in the cluster is given by $\epsilon_N = E_N/N$. We write all energies in units of k$_B$T. The total energy cost of forming all the clusters from dispersed monomers is given by:

$$U = \sum_{N=2}^{N=\infty} n_N \cdot E_N = \sum_{N=2}^{N=\infty} n_N \cdot N \cdot \epsilon_N. \tag{S5}$$

The summation starts at $N = 2$ since the monomer energy is $E_1 = 0$ by our definition of the reference energy. To obtain the energy density (in dimensionless units where the system size $M = V/v$), we divide Eq. S5 by the dimensionless system size, $M$, and insert $n_N$ from Eq. S3:

$$u = \frac{U}{M} = \sum_{N=2}^{N=\infty} N \cdot P_N \cdot \epsilon_N. \tag{S6}$$

Next, we derive the translational entropy density, which can be conveniently expressed as the sum of translational entropies of the monomers and of the clusters:

$$s = s_m + \sum_{N=2}^{N=\infty} s_N. \tag{S7}$$

With the assumption of a dilute solution ($P_N, \phi \ll 1$), the translational entropy of a cluster of size $N$ is given by

$$s_N = -P_N \ln P_N + P_N. \tag{S8}$$

The monomer entropy is

$$s_m = -\phi_m \ln \phi_m + \phi_m. \tag{S9}$$

The free energy density in units of k$_B$T is then given by

$$f = \sum_{N=2}^{N=\infty}(u_N - s_N) - s_m. \tag{S10}$$

Here $u_N = N \cdot P_N \cdot \epsilon_N$. Minimizing the free energy with respect to $P_N$, we find

$$\frac{\partial f}{\partial P_N} = N \cdot \epsilon_N + \ln P_N - N \ln \phi_m = 0. \tag{S11}$$

With the minimization of the monomer entropy performed using the chain rule

$$\frac{\partial s_m}{\partial P_N} = \frac{\partial s_m}{\partial \phi_m}\frac{\partial \phi_m}{\partial P_N}. \tag{S12}$$

At equilibrium, the cluster number concentration containing $N$ monomers satisfy

$$P_N = \exp[-N\epsilon_N + N \ln(\phi_m)]. \tag{S13}$$

We replace $\phi_m$ using Eq. S4 and rewrite Eq. S13 as a self-consistent equation for $P_N$

$$P_N = \exp[-N\epsilon_N + N \ln \phi] \cdot \left(1 - \sum_{j=2}^{j=\infty}\frac{j \cdot P_j}{\phi}\right)^N. \tag{S14}$$

## 2. Derivation in the grand canonical ensemble

We now consider an equivalent derivation of Eq. S14 for the case where the system is coupled to a protein monomer reservoir, with chemical potential $\mu$. In the actual system, this reservoir is the dilute solution of protein monomers themselves. The free energy density of the clusters is

$$f = \sum_{N=2}^{N=\infty}(u_N - s_N) + \mu\phi_m = \sum_{N=2}^{N=\infty}(u_N - s_N - \mu \cdot N \cdot P_N) + \mu\phi. \quad (S15)$$

Again, we minimize the free energy with respect to $P_N$ and find

$$\frac{\partial f}{\partial P_N} = N \cdot \epsilon_N + \ln P_N - \mu \cdot N = 0. \quad (S16)$$

The cluster number concentration is given by

$$P_N = \exp[-N(\epsilon_N - \mu)]. \quad (S17)$$

Therefore, we identify the chemical potential (written in units of k$_B$T) as $\mu = \ln \phi_m$, which is, in fact, the thermodynamic chemical potential (the derivative of the total free energy with respect to the number of monomers) for a monomer in a dilute solution.

### 3. The most common cluster size

Next, we derive an expression for the number of proteins in the most common cluster, $N^*$. A full solution involves solving the equation S14 and self-consistently deriving the distribution $P_N$. However, since this is analytically challenging, we use a saddle-point approximation – the clusters are assumed to be monodisperse, so the proteins can either be in dilute solution as monomers or in clusters with exactly $N = N^*$ monomers:

$$\sum_{j=2}^{j=\infty} j \cdot P_j \cong N^* \cdot P_{N^*}. \quad (S18)$$

The relation between the volume fraction of monomers and the protein volume fraction in the entire system is given by

$$\phi = \phi_m + N^* \cdot P_{N^*} \quad (S19)$$

With these simplifications, Eq. S14 can be written as:

$$\frac{N^* \cdot P_{N^*}}{\phi} = \exp[-N^*\epsilon_{N^*} + N^* \ln \phi + \ln N^* - \ln \phi] \cdot \left(1 - \frac{N^* \cdot P_{N^*}}{\phi}\right)^{N^*}. \quad (S20)$$

The assumption of a monodisperse distribution of clusters is most appropriate when the probability of finding a protein monomer in a cluster is maximal. That is, maximal volume concentration. Therefore, we maximize the product of the cluster number concentration $P_{N^*}$ and the number of proteins with respect to $N^*$. This means that we maximize the volume concentration:

$$N^*P_{N^*} = \max(N \cdot P_N). \tag{S21}$$

This must be solved for a specific model specifying the energy per monomer in a cluster of size $N$, $\epsilon_N$.

## 4. A specific model for cluster energy – amphiphilic layer

We consider two contributions to cluster formation energy – the core and the shell. The work associated with transferring a protein from a dilute solution to the cluster core, where it interacts mainly with other proteins rather than the aqueous solution, is its solubility energy, $\epsilon_B$. This includes the favorable protein-protein interaction energies that drive the cluster's formation, as well as the entropic cost associated with the reduction in the protein's conformational freedom within the denser cluster compared to a dilute solution. This does not yet include the translational entropy of the monomers, which we account for separately (described in section I.1). The total solubility energy per cluster is the product of the number of proteins in the cluster, $N$, and the solubility energy per monomer,

$$U_{core}(N) = N\epsilon_B. \tag{S22}$$

The interfacial energy per cluster in aqueous solution arises from the cluster "shell" (interfacial layer). Modeled as a spherical amphiphile layer with interfacial tension, $\gamma$, bending rigidity $\kappa$, saddle-splay rigidity, $\kappa_G$, and spontaneous curvature $J_s$ [1-3],

$$U_{shell}(R) = 4\pi\gamma R^2 - 8\pi\kappa J_s R + 8\pi\kappa + 4\pi\kappa_G. \tag{S23}$$

Since the constant contributions of saddle splay and bending cannot be distinguished in spherical geometry, we merge them into a single term with $\bar{\kappa} = \kappa + \frac{1}{2}\kappa_G$.

The total energy of forming a spherical cluster with $N$ monomers and radius $R$ (not including translational entropies of the cluster or the monomers in the solution from which the cluster is formed) is the sum of Eqs. S22 and S23. However, to solve Eq. S14 for the cluster number concentration $P_N$, we are interested in the energy as a function of $N$ alone. For simplicity, we therefore assume that the monomer volume, $v$, in a cluster is uniform and independent of the cluster size. Thus, the relation between $N$ and $R$ is

$$N = \frac{4\pi}{3}\frac{R^3}{v}. \tag{S24}$$

$v$ is the monolayer volume, which is assumed to be identical in all configurations. We insert Eq. S24 and find the sum of surface and core energies as a function of $N$

$$E_N = \epsilon_B \cdot N + 4\pi\gamma \left(\frac{3v}{4\pi}\right)^{\frac{2}{3}} \cdot N^{\frac{2}{3}} - 8\pi\kappa J_s \left(\frac{3v}{4\pi}\right)^{\frac{1}{3}} \cdot N^{\frac{1}{3}} + 8\pi\bar{\kappa}. \tag{S25}$$

And the energy per protein monomer

$$\epsilon_N = \frac{E_N}{N} = \epsilon_B + 4\pi\gamma \left(\frac{3v}{4\pi}\right)^{\frac{2}{3}} \cdot N^{-\frac{1}{3}} - 8\pi\kappa J_s \left(\frac{3v}{4\pi}\right)^{\frac{1}{3}} \cdot N^{-\frac{2}{3}} + 8\pi\bar{\kappa} \cdot N^{-1}. \tag{S26}$$

To clean up this expression, we define the prefactors of the various powers of $N$, with the numerical constants

$$\epsilon_A = 4\pi\gamma \left(\frac{3}{4\pi}v\right)^{\frac{2}{3}}, \quad \epsilon_R = -8\pi\kappa J_s \left(\frac{3}{4\pi}v\right)^{\frac{1}{3}}, \text{ and } \epsilon_T = 8\pi\bar{\kappa} \tag{S27}$$

So, the energy per cluster is given by

$$E_N = \epsilon_B N + \epsilon_A N^{\frac{2}{3}} + \epsilon_R N^{\frac{1}{3}} + \epsilon_T. \tag{S28}$$

The first term in Eq. S28 is the core solubility energy proportional to the cluster volume $N \sim R^3$, the second is the interfacial tension proportional to the cluster area $N^{\frac{2}{3}} \sim R^2$, the third is due to spontaneous curvature proportional to the cluster radius $N^{\frac{1}{3}} \sim R$, and the fourth is the 'topological term,' due to the bending energies, which, for spherical clusters, is independent of $N$ or the cluster radius $R$. The energy per monomer in a cluster of size $N$ is given by

$$\epsilon_N = \frac{E_N}{N} = \epsilon_B + \epsilon_A N^{-\frac{1}{3}} + \epsilon_R N^{-\frac{2}{3}} + \epsilon_T N^{-1}. \tag{S29}$$

Next, we derive the most common value of the cluster size $N \equiv N^*$ in two limits:

i. The formation energy of clusters is positive (unfavorable compared to the chemical potential of a monomer in solution), so the overall volume fraction of proteins in clusters is small compared with the total volume fraction of protein in the system ($\sum_{N=2}^{N=\infty} N \cdot P_N / \phi \ll 1$). In this regime, we expect transient oligomers (as fluctuations in the system) with $N^*$ of order unity.

ii.  The solubility energy is large (and negative relative to the chemical potential of a monomer in solution), and cluster formation is favorable. In this regime, the number of proteins in clusters is large, $N^* \gg 1$.

## 5. Classical self-assembly theory of small oligomers $N^* \sim 1$

We first consider the limit of small oligomers with the most common cluster number $N^*$ of order unity, and $\sum_{N=2}^{N=\infty} N \cdot P_N/\phi \ll 1$ (i.e., the total fraction of protein in the clusters is small). This is a simple case in which the solubility energy is either positive or negative but small (in units of k$_B$T). In this limit, we can approximate

$$\left(1 - \frac{NP_N}{\phi}\right)^N \cong 1 - N\frac{N \cdot P_N}{\phi}, \quad (S30)$$

We solve for $P_N$ from Eq. S20,

$$P_N = \frac{\exp[-N\epsilon_N + N\ln\phi]}{1 + \frac{N^2}{\phi}\cdot\exp[-N\epsilon_N + N\ln\phi]}. \quad (S31)$$

Since the solubility energy is close to zero, $\frac{N^2}{\phi}e^{-N\epsilon_B + N\ln\phi} \ll 1$ we find,

$$P_N \cong \exp[-N\epsilon_N + N\ln\phi]. \quad (S32)$$

The protein volume fraction roughly equals the monomers since few monomers are in clusters. This is the classical self-assembly result when $\phi_m \approx \phi$ so that chemical potential $\mu = \ln\phi_m \approx \ln\phi$. Considering the shell and core terms separately is not needed since the clusters in this case are small, and almost all protein molecules are in contact with the aqueous environment. Therefore, we consider only the solubility energy in Eq. S29 and find the cluster number concentration,

$$P_N = \exp[-\epsilon_B N + N\ln\phi], \quad (S33)$$

And volume fraction

$$NP_N = \exp[-\epsilon_B N + N\ln\phi + \ln N]. \quad (S34)$$

Maximizing the argument in Eq. S34 with respect to $N$ to find $N^*$ such $NP_N$ is maximal

$$N^* = \frac{1}{\epsilon_B - \ln\phi}. \quad (S35)$$

The oligomers' volume fraction is found by inserting Eq. S35 into Eq. S32,

$$N^* P_{N^*} = \frac{e^{-1}}{\epsilon_B - \ln \phi} = \frac{e^{-1}}{N^*}. \quad (S36)$$

We note that Eqs. S35 and S36 are valid only if $\epsilon_B - \ln \phi > 0$ but $\frac{1}{\epsilon_B - \ln \phi} \sim 1$ which is satisfied for either large and positive values of $\epsilon_B$ or small enough values of $\phi$ even if $\epsilon_B < 0$. When $\epsilon_B$ approaches $\ln \phi$ the denominator of Eq. S34 goes to zero so that the most common number of monomers in a cluster, $N^*$, and the total volume fraction becomes large, and our initial assumption of $N^*$ order unity does not hold. The case of large $N^*$, relevant to the experiments we discuss, is treated in the next section.

**6. Self-assembly of large clusters $N^* \gg 1$**

When the solubility energy is negative and approaches the chemical potential, $\epsilon_B \to \mu$, the number of proteins in the clusters grows. The size distribution is determined by the balance between entropy, which favors a uniform distribution of clusters of all sizes, and their interfacial energy, which favors the merger of the clusters to reduce the interfacial free energy cost per monomer. If the interfacial energy is the order of a few k$_B$T, we expect to find a wide distribution of cluster sizes with $N^* \gg 1$. We consider two limits in the following:

i. The system is well below the critical aggregation concentration (CAC), meaning that most proteins are in monomers in a dilute solution and that the total volume fraction of clusters is very low ($\sum_{N=2}^{N=\infty} N \cdot P_N / \phi \ll 1$). However, unlike the situation in the previous section, the combination of $\epsilon_B \to \ln(\phi)$ and a small interfacial energy result in large clusters ($N^* \gg 1$).

ii. The system is well above the CAC, and most of the proteins in the system are in clusters ($\sum_{N=2}^{N=\infty} N \cdot P_N / \phi \sim 1$) and the number of monomers is relatively small ($\phi_m \ll \phi$). The growth of the clusters is then limited by the energetic cost of their curvature, similar to microemulsions.

(i) *Below the CAC*
To obtain an expression for $P_N$ in this limit, we again perform a saddle-point approximation where we consider a monodisperse cluster distribution and search for $N = N^*$ such that the maximal number of proteins is in these clusters. To do so, we rewrite Eq. S20 as

$$\left(\frac{N \cdot P_N}{\phi}\right)^{\frac{1}{N}} = \exp\left[-\epsilon_N + \ln\phi - \frac{\ln\phi}{N} + \frac{\ln N}{N}\right] \cdot \left(1 - \frac{N \cdot P_N}{\phi}\right). \qquad (S37)$$

In the limit $N \gg 1$ and $N \cdot \frac{P_N}{\phi} \ll 1$ we can expand

$$\left(\frac{N \cdot P_N}{\phi}\right)^{\frac{1}{N}} \sim 1 + \frac{\ln\left(\frac{N \cdot P_N}{\phi}\right)}{N} \qquad (S38)$$

And Eq. S37 is approximated as

$$1 + \frac{\ln\left(\frac{N \cdot P_N}{\phi}\right)}{N} = \exp\left[-\epsilon_N + \ln\phi - \frac{\ln\phi}{N} + \frac{\ln N}{N}\right] \cdot \left(1 - \frac{N \cdot P_N}{\phi}\right). \qquad (S39)$$

We neglect the second term on the left-hand side and solve for $N \cdot P_N$,

$$N \cdot P_N = \phi - \exp\left[\epsilon_N + \frac{\ln\phi}{N} - \frac{\ln N}{N}\right]. \qquad (S40)$$

To find the maximal number of proteins in clusters, $N^* P_{N^*}$, we now minimize the exponent in Eq. S40 with respect to $N$. For the generic energy per monomer introduced in Eq. S29, the argument in the exponent is

$$\epsilon_B + \epsilon_A N^{-\frac{1}{3}} + \epsilon_R N^{-\frac{2}{3}} + \epsilon_T N^{-1} + \frac{\ln\phi}{N} - \frac{\ln N}{N}. \qquad (S41)$$

The extremum of this is found by solving

$$\frac{-\frac{1}{3}\epsilon_A N^{*\frac{2}{3}} - \frac{2}{3}\epsilon_R N^{*\frac{1}{3}} - \epsilon_T - \ln\phi - 1 + \ln N^*}{N^{*2}} = 0. \qquad (S42)$$

Eq. S42 has two solutions - finite $N^*$ and $N^* \to \infty$, respectively, representing a finite size (but large) most common cluster and macroscopic phase separation. We note that $\phi_C = \sum N P_N < \phi$ (Eq. S4) so if $N^* \to \infty$ then $P_{N^*} \to 0$ and the product $N^* \cdot P_{N^*}$ is always finite. A finite-size cluster solution is stable if the volume fraction obeys:

$$N^* P_{N^*} > \lim_{N \to \infty} (N \cdot P_N). \qquad (S43)$$

The right-hand side of Eq. S42 becomes:

$$\lim_{N \to \infty} (N \cdot P_N) = \phi - \exp[\epsilon_B]. \qquad (S44)$$

In this regime ($N \gg 1$, below the CAC) $\epsilon_B$ is negative so $\lim_{N \to \infty}(N \cdot P_N) \sim \phi$. To calculate the left-hand side of Eq. S43, we consider the simple case where the topological term, $\epsilon_T$ (due to the bending energies), dominates the interfacial energy, as discussed in the main text ($\epsilon_A N^{\frac{2}{3}} + \epsilon_R N^{\frac{1}{3}} \ll \epsilon_T$). In such case, we find $N^* P_{N^*}$ by solving Eq. S42 for finite $N^*$

$$\ln N^* = \epsilon_T + 1 + \ln \phi. \tag{S45}$$

Inserting this into Eq. S40 we find

$$N^* P_{N^*} = \phi - \exp[\epsilon_B - N^{*-1}] > \lim_{N \to \infty}(N \cdot P_N). \tag{S46}$$

That is, finite clusters with no surface tension and no spontaneous curvature are always favored over LLPS. However, for finite interfacial tension or spontaneous curvature, there can be a transition from a phase with finite clusters to LLPS. This can be shown in the limit of $\epsilon_A N^{\frac{2}{3}}, \epsilon_R N^{\frac{1}{3}} \gg \epsilon_T$. However, this is beyond the scope of the present work, as this limit is not relevant to the phases with clusters that we study here.

To compare our model to experimental data, we rewrite Eq. S45 in terms of cluster radius and protein molar concentration $C$ using the relations

$$\phi = vC \quad \text{and} \quad N^* = \frac{4\pi}{3} \frac{R^{*3}}{v}. \tag{S47}$$

Eq. S45 can be re-written as

$$\ln \frac{R^{*3}}{v} = \epsilon_T + 1 - \ln \frac{4\pi}{3} + \ln(C \cdot v) \tag{S48}$$

Noting that $\epsilon_T = 8\pi \bar{\kappa}/k_B T$ (Eq. S27).

*(ii) Above the CAC*

Significantly above the CAC, the number of monomers changes very slowly as the total protein number increases [2]. We denote the CAC protein monomer volume fraction as $\phi_m^{CAC}$ and $\Delta \phi$ as the increase in protein monomer concentration above the CAC, which is small relative to $\phi_m^{CAC}$ ($\phi_m^{CAC} \gg \Delta \phi_m$),

$$\phi_m = \phi_m^{CAC} + \Delta \phi_m. \tag{S49}$$

With that, Eq. S13 is written as

$$NP_N = \exp\left[-N\epsilon_N + N\ln(\phi_m^{CAC}) + N\ln\left(1 + \frac{\Delta\phi_m}{\phi_m^{CAC}}\right) + \ln N\right]. \quad (S50)$$

Since $\phi_m$ changes weakly with $\phi$ above the CAC, we approximate $\frac{\Delta\phi_m}{\phi_m^{CAC}} \ll 1$ and $\ln\left(1 + \frac{\Delta\phi_m}{\phi_m^{CAC}}\right) \sim \frac{\Delta\phi_m}{\phi_m^{CAC}}$. So, Eq. S50 is approximated to a monodisperse cluster distribution of clusters of size $N$,

$$NP_N = \exp\left[-N\epsilon_N + N\ln(\phi_m^{CAC}) + N\frac{\Delta\phi_m}{\phi_m^{CAC}} + \ln N\right]. \quad (S51)$$

We maximize Eq. S51 by finding the maximum of the argument in the exponential with respect to $N$:

$$-\epsilon_B - \frac{2}{3}\epsilon_A N^{*-\frac{1}{3}} - \frac{1}{3}\epsilon_R N^{*-\frac{2}{3}} + \ln(\phi_m^{CAC}) + \frac{\Delta\phi_m}{\phi_m^{CAC}} + \frac{1}{N^*} = 0 \quad (S52)$$

When solubility energy dominates, $(|\epsilon_B| \gg \epsilon_A N^{*-\frac{1}{3}}, \epsilon_R N^{*-\frac{2}{3}})$ we find

$$\frac{1}{N^*} = \epsilon_B - \ln(\phi_m^{CAC}) - \frac{\Delta\phi_m}{\phi_m^{CAC}}. \quad (S53)$$

We define the CAC as the concentration of protein where $\Delta\phi_m = 0$, so that the cluster size at the CAC is given by (Eq. S53):

$$N_{CAC} = \frac{1}{\epsilon_B - \ln(\phi_m^{CAC})}. \quad (S54)$$

And the (most common) cluster volume fraction at the CAC by inserting $N_{CAC}$ and $\Delta\phi_m = 0$ to Eq. S51

$$N_{CAC}P_{N_{CAC}} = N_{CAC}\exp[-1 - \epsilon_T]. \quad (S55)$$

The cluster volume fraction far above the CAC is found by inserting $N^*$ (Eq. S53) into Eq. S51.

$$N^*P_{N^*} = \frac{N_{CAC}P_{N_{CAC}}}{1 - N_{CAC}\frac{\Delta\phi_m}{\phi_m^{CAC}}}. \quad (S56)$$

In the limit of $N_{CAC}\frac{\Delta\phi_m}{\phi_m^{CAC}} \ll 1$ we have

$$N^*P_{N^*} = N_{CAC}P_{N_{CAC}}\left(1 + N_{CAC}\frac{\Delta\phi_m}{\phi_m^{CAC}}\right). \quad (S57)$$

With the constraint of a fixed number of proteins, we find

$$\phi = \phi_m + N^* P_{N^*} = \phi_m^{CAC} + N_{CAC} P_{N_{CAC}} + \left(N_{CAC} P_{N_{CAC}} \cdot \frac{N_{CAC}}{\phi_m^{CAC}} + 1\right) \Delta \phi_m. \quad (S58)$$

The CAC is defined as the point beyond which most of the newly added protein forms new clusters. This definition is somewhat arbitrary and is defined as

$$N_{CAC} P_{N_{CAC}} = f_{CAC} \phi_{CAC} \quad (S59)$$

With $0 < f_{CAC} < 1$. At the CAC, we have $\phi_{CAC} = f_{CAC}\phi_{CAC} + \phi_m^{CAC}$. So, the monomer concentration at the CAC is

$$\phi_m^{CAC} = \phi_{CAC}(1 - f_{CAC}) \quad (S60)$$

If $f = \frac{1}{2}$ as defined in Ref. [2], then $\phi_m^{CAC} = N_{CAC} P_{N_{CAC}} = \frac{1}{2}\phi_{CAC}$. With this definition, we simplify Eq. S58

$$\phi = \phi_{CAC} + \left(\frac{f_{CAC}}{1 - f_{CAC}} \cdot N_{CAC} + 1\right) \Delta \phi_m. \quad (S61)$$

So, the increase in monomer fraction significantly above the CAC is

$$\Delta \phi_m = \frac{1 - f_{CAC}}{f_{CAC}} \frac{\phi - \phi_{CAC}}{N_{CAC}}. \quad (S62)$$

Here we used $\frac{f_{CAC}}{1-f_{CAC}} \cdot N_{CAC} \gg 1$. By inserting Eqs. S62, S59, and S60 to Eq. S57, we find the change in cluster volume fraction above the CAC

$$\Delta(N^* P_{N^*}) = N^* P_{N^*} - N_{CAC} P_{N_{CAC}} \cong \phi - \phi_{CAC} \quad (S63)$$

Finally, we can rewrite Eq. S53 using Eqs. S54 and S62

$$\frac{N_{CAC}}{N^*} = 1 + \frac{1}{f_{CAC}} - \frac{1}{f_{CAC}} \frac{\phi}{\phi_{CAC}} \quad (S64)$$

To fit the experimental data, we rewrite Eq. S65 by inserting Eq. S47

$$\frac{R_{CAC}^3}{R^{*3}} = 1 + \frac{1}{f_{CAC}} - \frac{1}{f_{CAC}} \frac{C}{C_{CAC}} \quad (S65)$$

$C$ is the protein molar concentration, $C_{CAC}$ is the molar concentration at the CAC, and $\Delta C$ the increase beyond the CAC.

## II. Supplementary Figures and Tables – FUS and CPEB4
## Supplementary Figure S1 – Protein structure prediction using AlphaFold

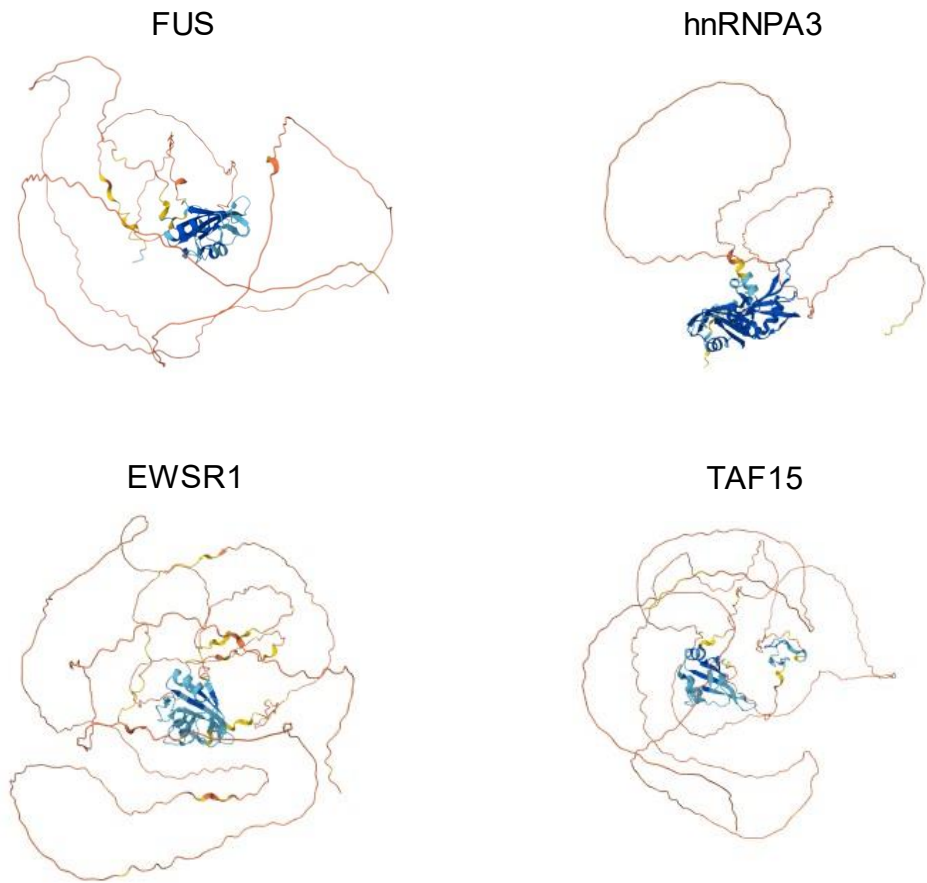

The predicted structures of the proteins that are presented in this work. All structures are obtained from the AlphaFold [4] Protein Structure Database https://alphafold.ebi.ac.uk/. UniProt IDs: FUS-P35637, hnRNPA3- P51991, EWSR1 – Q01844, and TAF15- Q92804. The CPEB4$_{NTD}$ structure was not available.

**Supplementary Fig. S2 – CPEB4 and FUS (DLS and NTA) in linear-linear scale**

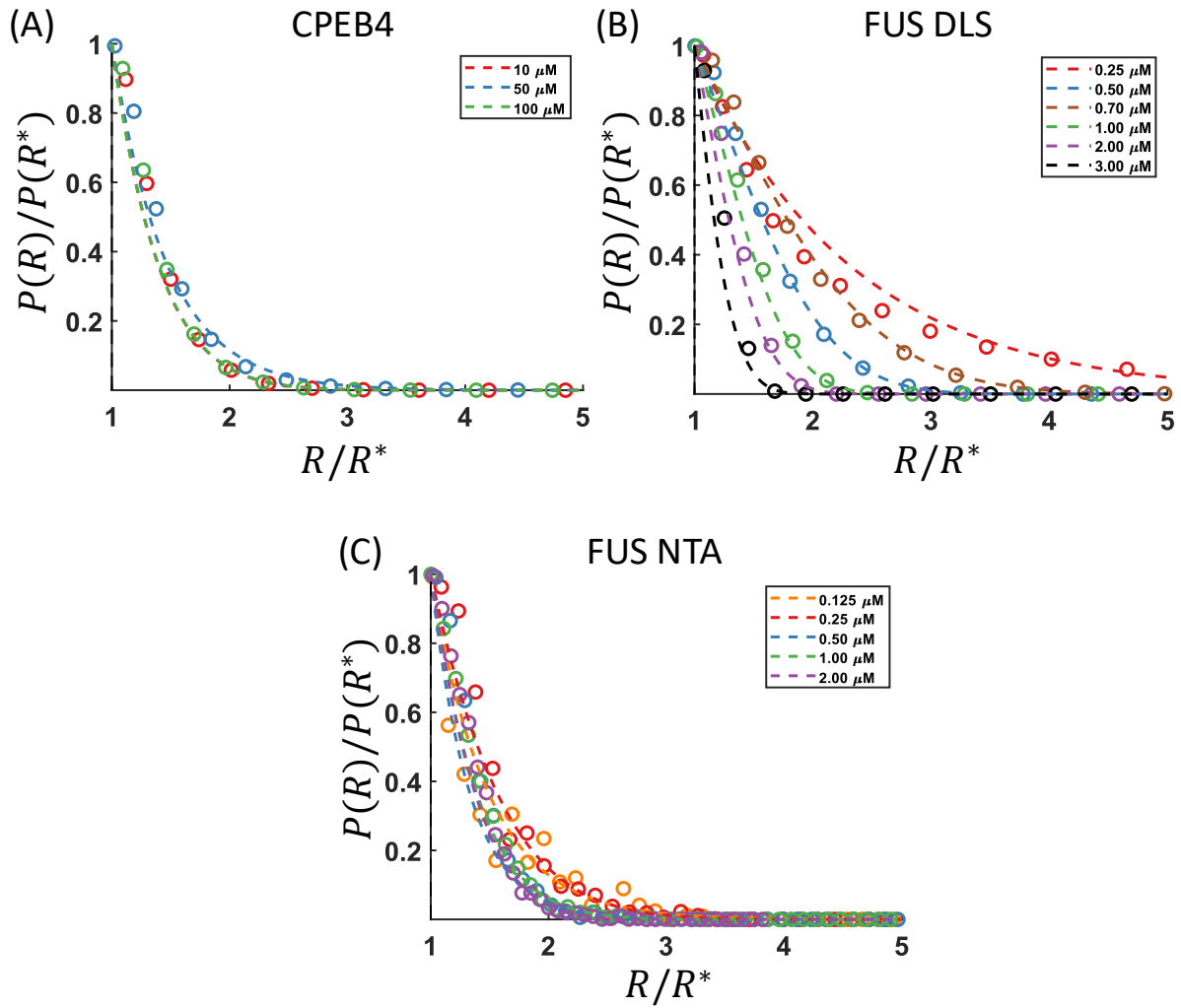

Normalized number distribution in linear-linear scale. Open circles – experimental data, dashed lines – theoretical fit to Eq. 16 (main text). The protein concentration is indicated in the legends. (A) CPEB4 at 277 °K, 25 mM Tris, pH 8, and 100 mM NaCl corresponding to Fig. 3A. (B and C) FUS at 298 °K, 20 mM Tris, pH 7.4, and 100 mM KCl. (B) DLS data corresponding to Fig. 3 B and (C) NTA data corresponding to Fig. 4.

**Supplementary Table S1 – FUS and CPEB4 Hydrophobicity index**

|        | Total Hydrophobic Blob | Total Hydrophilic Blob | Averaged Hydrophobic Blob | Averaged Hydrophilic Blob |
|---|---|---|---|---|
| **FUS** | -106 | -308, -280 | -1.21 | -1.23, -1.45 |
| **CPEB4$_{NTD}$** | -55 | -323 | -1.17 | -0.80 |

Kyte-Doolittle hydrophobicity index [5] of FUS and CPEB4$_{NTD}$. More negative values represent greater hydrophobicity. For FUS, the hydrophobic blob is approximated here to be the RNA binding domain, and the hydrophilic blob is approximated here by the intrinsically disordered part of the protein. The total is the sum of the hydrophobicity index for all the amino acids (AA), while the last two columns are the index per AA. FUS has two disordered domains, which we consider as the hydrophobic blobs – one between AA 1-213 and the second AA 302-528. The hydrophilic blob is AA 214-301. The CPEB4 protein considered here contains only the intrinsically disordered domain (AA 1-448). In previous work (SI of [3], Fig. S14), we found that the hydrophobic blob is AA 1-300 and the hydrophilic blob is AA 301-448. The ratio of hydrophobic to hydrophilic AA is about 3:1 for FUS and about 6:1 for CPEB4 within this crude estimate. This would imply that the preferred packing of the amphiphilic layer should be with the hydrophobic blobs on the "outside" of the cluster, which is opposite to the packing required in an aqueous solution. This may be the reason that the spontaneous (preferred) curvature energies in the Tables below are positive, corresponding to preferred packing with the hydrophobic blobs on the "outside".

**Supplementary Table S2 – Fit results for CPEB4**

| Concentration [µM] | 10 | 50 | 100 |
|---|---|---|---|
| $R^*$ [nm] | 17 | 21 | 23 |
| Curvature tendency energy - $A_1$ | 1.9 | 2.14 | 1.88 |
| Interfacial tension energy – $A_2$ | 0.26 | 0 | 0.29 |
| Core energy – $A_3$ | 0 | 0 | 0 |
| Interfacial tension $\gamma$ [µN/m] | 0.30 | 0 | 0.17 |
| $J_s \kappa_B$ [fN] | -18 | -17 | -13 |
| $\chi^2_{min}$ | 0.15 | 0.87 | 0.19 |

CPEB4 fitting result, corresponding to Fig. 3 A. The dimensionless parameters $A_1$, $A_2$, and $A_3$ are found by a least squares minimization procedure (Eq. 18). The interfacial tension, $\gamma$, is derived using Eq. 17 (b) and $J_s\kappa$ using Eq. 5 (a). We are aware that the fits for 50 µM gave a vanishing tension ($A_2$); however, the value of $A_2$ has a much smaller impact on the normalized cluster size distribution compared to the curvature tendency and is thus more sensitive to noise. Therefore, the fitted values of $A_2$ and $\gamma$ should be considered as order of magnitude estimates.

**Supplementary Table S3 – Fit results for FUS DLS**

| Concentration [µM] | 0.25 | 0.5 | 0.7 | 1 | 2 | 3 |
|---|---|---|---|---|---|---|
| $R^*$ [nm] | 49 | 147 | 149 | 225 | 336 | 381 |
| Curvature tendency energy - $A_1$ | 0.76 | 0.19 | 0.07 | 0 | 0 | 0 |
| Interfacial tension energy – $A_2$ | 0 | 0.2 | 0.29 | 0 | 0 | 0 |
| Core energy – $A_3$ | 0 | 0.1 | 0 | 0.39 | 0.6 | 1.18 |
| Interfacial tension $\gamma$ [nN/m] | 0 | 3.1 | 4.3 | 0 | 0 | 0 |
| $J_s \kappa_B$ [fN] | -2.5 | -0.21 | -0.07 | 0 | 0 | 0 |
| $\chi^2_{min}$ | 0.37 | 0.02 | 0.04 | 0.06 | 0.16 | 0.69 |

FUS DLS fitting results corresponding to Fig. 3 B. The dimensionless parameters $A_1$, $A_2$, and $A_3$ are found by a least-squares minimization procedure (Eq. 18). The interfacial tension, $\gamma$, is derived using Eq. 17 (b) and $J_s\kappa$ using Eq. 17 (a).

**Supplementary Table S4– Fit results for FUS NTA**

| Concentration [μM] | 0.125 | 0.25 | 0.5 | 1 | 2 |
|---|---|---|---|---|---|
| $R^*$ [nm] | 37 | 34 | 41 | 47 | 66 |
| Curvature tendency energy - $A_1$ | 2.04 | 0.99 | 3.02 | 1.69 | 0.79 |
| Interfacial tension energy – $A_2$ | 0 | 0.3 | 0 | 0.37 | 0.80 |
| Core energy – $A_3$ | 0 | 0 | 0 | 0 | 0 |
| Interfacial tension $\gamma$ [nN/m] | 0 | 84 | 0 | 54 | 60 |
| $J_s \kappa_B$ [fN] | -9.10 | -4.75 | -12.2 | -5.9 | -1.97 |
| $\chi^2_{min}$ | 5.42 | 2.84 | 4.42 | 1.75 | 7.42 |

FUS NTA fitting result corresponding to Fig. 4. The dimensionless parameters $A_1$, $A_2$, and $A_3$ are found by a least squares minimization procedure (Eq. 18). The interfacial tension, $\gamma$, is derived using Eq. 17 (b) and $J_s \kappa$ using Eq. 17 (a).

## III. Supplementary Note S2: Size distribution of EWSR1, TAF15 and hnRNPA3

EWSR1, TAF15, and hnRNPA3 are proteins of the FET family that also form clusters under sub-saturated conditions [6]. Similar to FUS, they have a highly ordered RNA-binding domain and a long intrinsically disordered tail (Fig. S2). To verify whether our theoretical model accurately describes the cluster size distribution formed by these proteins, we followed the same procedure described in the main text to fit the experimental DLS intensity data previously published [6]. To account for the difference between the number distribution used in Eq. 16 and the intensity distribution, we multiply the fractions by $R^6$ to account for the Raleigh scattering. The normalized intensity profile is given by,

$$\left(\frac{R}{R^*}\right)^{-6} \frac{I(R)}{I(R^*)} = \exp\left[-A_3\left(\frac{R^3}{R^{*3}} - 1\right) - A_2\left(\frac{R^2}{R^{*2}} - 1\right) - A_1\left(\frac{R}{R^*} - 1\right)\right]. \tag{S66}$$

The least-square minimization process is done similarly to FUS and CPEB4, but now includes the $\left(\frac{R}{R^*}\right)^6$ factor. The results are presented in Fig. S4. The detailed fitting results are presented in Tables S5 (EWSR1), S6 (TAF15), and S7 (hnRNPA3).

Similar to the low concentration FUS and CPEB4, we found the cluster size distribution is dominated by the curvature tendency, which is 3-5 k$_B$T per cluster. $R^*$ increases significantly with protein concentration, as in FUS. However, in contrast, the tension contribution is negligible compared to the curvature tendency energy. To conclude, the cluster size distributions of EWSR1, TAF15, and hnRNPA3 are qualitatively similar to FUS and can be explained by our theoretical model.

**Supplementary Figure S4 – Fit of theory to experiment**

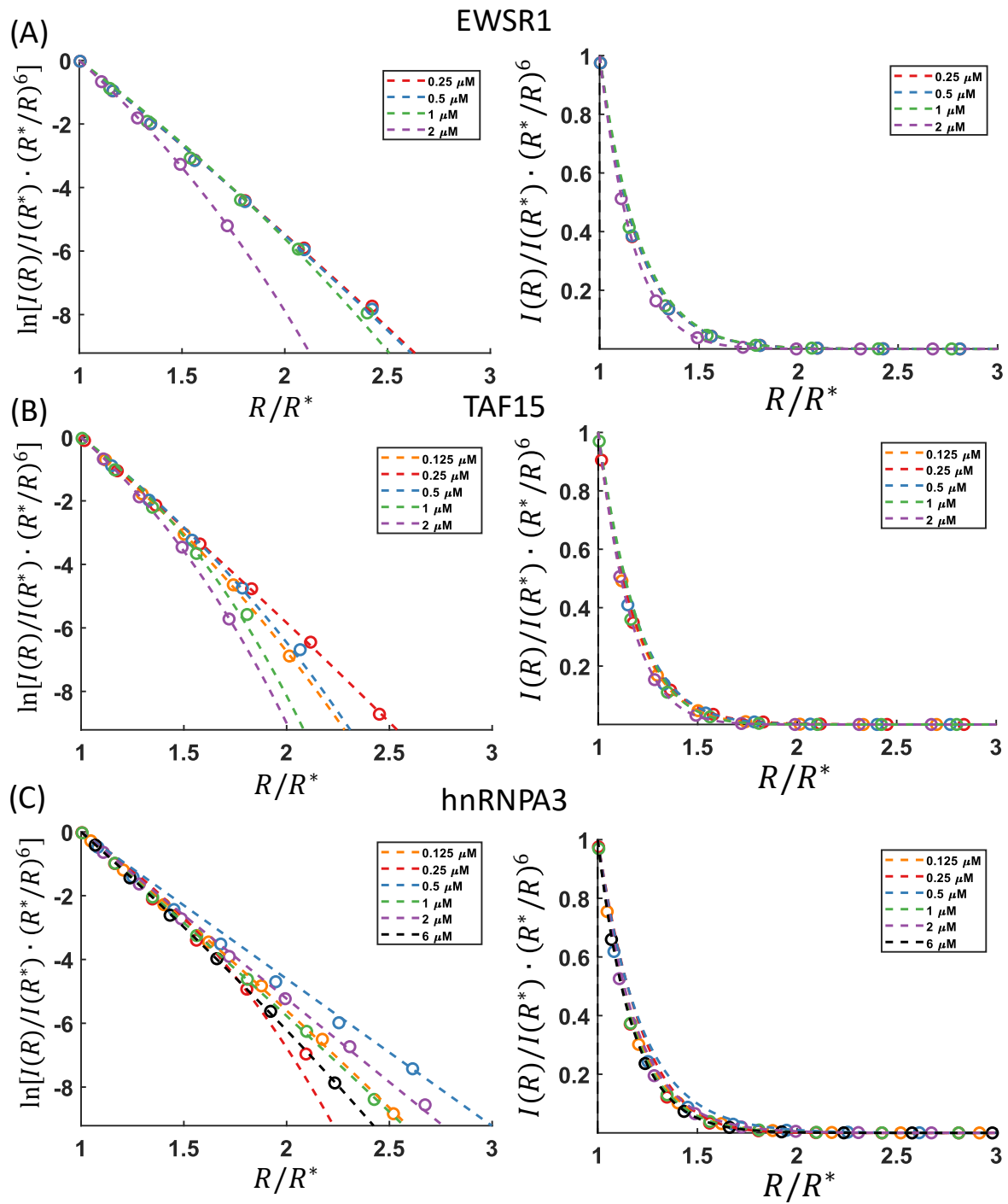

Normalized intensity profiles: (A) EWSR1, (B) TAF15, and (C) hnRNPA3 DLS. Data taken from [6]. All measurements were performed at 298 °K, 20 mM Tris, pH 7.4, and 100 mM KCl. The proteins are SNAP-tagged, as explained in [6]. Open circles – experimental data, dashed lines – theoretical fit to Eq. S66. The protein concentration is indicated in the legends. The results of the fitting are summarized in Table S5 for EWSR1, Table S6 for TAF15, and Table S7 for hnRNPA3. LLPS occurs at concentrations of ~6 µM for hnRNPA3, 2 µM for EWSR1, and 2 µM for TAF15.

**Supplementary Table S5 – Fit results for EWSR1**

| Concentration [µM] | 0.25 | 0.5 | 1 | 2 |
|---|---|---|---|---|
| $R^*$ [nm] | 82 | 82 | 231 | 372 |
| Curvature tendency energy - $A_1$ | 4.96 | 4.7 | 4.3 | 4.16 |
| Interfacial tension energy – $A_2$ | 0.1 | 0.22 | 0.01 | 0.1 |
| Core energy – $A_3$ | 0.03 | 0.02 | 0.18 | 0.49 |
| Interfacial tension $\gamma$ [nN/m] | 4.94 | 10.8 | 0.06 | 0.24 |
| $J_s \kappa_B$ [fN] | -10 | -9.5 | -3.1 | -18 |
| $\chi^2_{min}$ | 0.12 | 0.12 | 0.44 | 0.002 |

EWSR1 fitting result, corresponding to Fig. S4 A. The dimensionless parameters $A_1$, $A_2$, and $A_3$ are found by a least squares minimization procedure with the additional factor $(R^*/R)^6$ (Eq. S66) that corrects for the Rayleigh scattering intensity. The interfacial tension, $\gamma$, is derived using Eq. 17 (b) and $J_s\kappa$ using Eq.17 (a).

**Supplementary Table S6 – Fit results for TAF15**

| Concentration [µM] | 0.125 | 0.25 | 0.5 | 1 | 2 |
|---|---|---|---|---|---|
| $R^*$ [nm] | 132 | 125 | 231 | 264 | 372 |
| Curvature tendency energy - $A_1$ | 4.6 | 5.0 | 3.8 | 2.2 | 3.2 |
| Interfacial tension energy – $A_2$ | 0 | 0.28 | 0.10 | 0 | 0.07 |
| Core energy – $A_3$ | 0.30 | 0 | 0.33 | 0.85 | 0.80 |
| Interfacial tension $\gamma$ [nN/m] | 0 | 5.9 | 0.6 | 0 | 0.2 |
| $J_s \kappa_B$ [fN] | -5.8 | -6.6 | -2.7 | -1.4 | -1.4 |
| $\chi^2_{min}$ | 0.01 | 0.03 | 0.20 | 0.23 | 0.005 |

TAF15 fitting result, corresponding to Fig. S4 B. The dimensionless parameters $A_1$, $A_2$, and $A_3$ are found by a least squares minimization procedure with the additional factor $(R^*/R)^6$ (Eq. S66) correcting for the Rayleigh scattering intensity. The interfacial tension, $\gamma$, is derived using Eq. 17 (b) and $J_s\kappa$ using Eq.17 (a).

**Supplementary Table S7 – Fit results for hnRNPA3**

| Concentration [µM] | 0.125 | 0.25 | 0.5 | 1 | 2 | 6 |
|---|---|---|---|---|---|---|
| $R^*$ [nm] | 51 | 82 | 88 | 109 | 154 | 248 |
| Curvature tendency energy - $A_1$ | 4.8 | 2.5 | 4.6 | 5.6 | 5.2 | 5.4 |
| Interfacial tension energy – $A_2$ | 0.22 | 0 | 0 | 0 | 0 | 0 |
| Core energy – $A_3$ | 0.2 | 0.6 | 0 | 0.03 | 0 | 0.11 |
| Interfacial tension $\gamma$ [nN/m] | 29 | 0 | 0 | 0 | 0 | 0 |
| $J_s \kappa_B$ [fN] | -15.5 | -5.0 | -8.6 | -8.4 | -5.6 | -3.6 |
| $\chi^2_{min}$ | 0.05 | 0.97 | 0.6 | 0.02 | 0.14 | 0.01 |

hnRNPA3 fitting result, corresponding to Fig. S4 C. The dimensionless parameters $A_1$, $A_2$, and $A_3$ are found by a least squares minimization procedure with the additional factor $(R^*/R)^6$ (Eq. S66) that corrects for the Rayleigh scattering intensity. The interfacial tension, $\gamma$, is derived using Eq. 17 (b) and $J_s \kappa$ using Eq.17 (a).

## IV. Supplementary Note S3: Limitations of fitting scattering in the range $R < R^*$

The DLS and NTA measurements also include data points for values of the cluster radius $R$ smaller than the most probable cluster radius, $R^*$. However, we did not include the range $R < R^*$ in the fits presented in Figures 3 and 4 (and their corresponding figures in the Supporting Information) due to theoretical limitations of our model and, more significantly, the substantial bias inherent to experimental measurements within this range.

Our theoretical framework predicts the size distribution for clusters with $R > R^*$ (Eq. 16, main text), and remains valid as long as the cluster size is significantly greater than the characteristic molecular diameter related to the protein size. This condition is met by the clusters in FUS and the other FUS-like proteins, as measured via DLS; those clusters have radii which are typically hundreds of nanometers so that they substantially exceed the dimensions of individual protein molecules. However, DLS measurements are particularly prone to biases caused by the presence of large clusters in solution. As an illustrative example, we present the fitting around the distribution peak (for both $R > R^*$ and $R < R^*$ for the FUS DLS data (corresponding to Fig. 3 B in the main text):

**Supplementary Figure S5 – Fit of theory to experiment for the region of R near R***

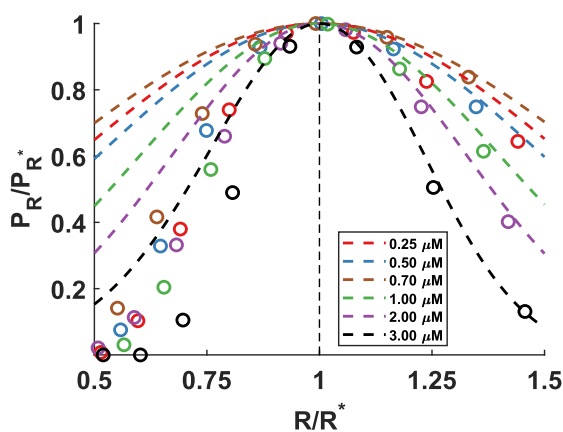

As clearly demonstrated, the fits agree well with experimental data in the range $R > R^*$ but deviate notably in the range $R < R^*$, especially far from the peak. Notably, the smallest observable clusters appear only at approximately $R \approx 0.5\, R^*$, independent of protein concentration and the scattering of smaller cluster is not observed. For FUS clusters, $R^*$ spans the range from roughly 50 nm at 0.25 µM to 350 nm at 3 µM. However, according to NTA measurements (Fig. 4, main text), small clusters indeed exist, but these fall below the detection limits of DLS. This shows that the DLS data reliability diminishes substantially in the range $R < R^*$.

Clusters observed in the measurements of FUS NTA and CPEB4 DLS within the range $R < R^*$ are extremely small (Figs. 4 and 3A, respectively), with sizes approaching that of the protein molecules. Such small clusters are inadequately captured by our current theoretical approach, which assumes the clusters are much larger than the molecular size. Generalizing this to smaller radii would require incorporating higher-curvature corrections, which will introduce more theoretical parameters and are beyond the scope of this study.

In summary, the FUS DLS data are strongly influenced by the experimental biases that give more weight to larger clusters, while the CPEB4 DLS and FUS NTA data are not well described by our theory for radii smaller than $R^*$. We do note that the data points for values of $R$ near $R^*$ are indeed adequately described by the theory. However, these points have a minimal influence on overall fitting outcomes due to normalization constraints, since the fits are anchored at the peak. Therefore, to maintain clarity and focus on the primary insights of our study, we excluded the region $R < R^*$ from our fitting procedure.

### V. Supplementary References